\documentclass[a4paper,fleqn,usenatbib,useAMS]{mnras}

\usepackage{graphicx}	
\usepackage{amsmath}	
\usepackage{amssymb}	
\usepackage{multicol}        
\usepackage{bm}		
\usepackage{multirow}
\usepackage[usenames,dvipsnames]{xcolor}
\definecolor{sepia}{HTML}{7f462c }
\definecolor{darkgreen}{HTML}{22aa22 }

\newcommand{\Msun}{\ensuremath{M_{\odot}}}

\newcommand{\mypath}[1]{./#1}

\newcommand{\corrd}[1]{{\color{darkgreen} #1}}
\newcommand{\corr}[1]{{\color{red} #1}}    
\newcommand{\nuovo}[1]{{\color{red} #1}}    

\renewcommand{\corrd}[1]{{  #1}}
\renewcommand{\corr}[1]{{ #1}}    
\renewcommand{\nuovo}[1]{{ #1}}

\usepackage[T1]{fontenc}
\usepackage{ae,aecompl}

\title[$c_{200}\left(M_{200}, z, \rm{fossilness}\right)$]{Dependency of halo concentration on mass, redshift and fossilness in Magneticum hydrodynamic simulations}

\author[A. Ragagnin]{Antonio Ragagnin$^{1,2,3}$, Klaus Dolag$^{3,4}$, 
Lauro Moscardini $^{5,6,7}$, \newauthor Andrea Biviano$^{8}$, Mauro D'Onofrio$^{9}$
\\
$^{1}$ Leibniz-Rechenzentrum (LRZ), Boltzmannstrasse 1, 85748 Garching, Germany\\
$^{2}$ Excellence Cluster Universe, Boltzmannstrasse 2, 85748 Garching, Germany\\
$^{3}$ Universit\"ats-Sternwarte, Fakult\"at f\"ur Physik, Ludwig-Maximilians Universit\"at M\"unchen, Scheinerstrasse 1, 81679 M\"unchen, Germany\\
$^{4}$ Max-Planck-Institut f\"ur Astrophysik, Karl-Schwarzschild Strasse 1, 85748 Garching bei M\"unchen, Germany\\
$^{5}$ Dipartimento di Fisica e Astronomia, Alma Mater Studiorum - Universit\`a di Bologna, via Piero Gobetti 93/2, 40129 Bologna, Italy\\
$^{6}$ INAF - Osservatorio di Astrofisica e Scienza dello Spazio di Bologna, via Piero Gobetti 93/3, 40129 Bologna, Italy\\ 
$^{7}$ INFN, Sezione di Bologna, viale Berti Pichat 6/2, 40127 Bologna, Italy\\
$^{8}$ INAF - Osservatorio Astronomico di Trieste, via G.B. Tiepolo 11, 34143 Trieste, Italy\\
$^{9}$ Dipartimento di Fisica e Astronomia G. Galilei, Universit\`a di Padova, Vicolo Osservatorio 3, 35122 Padova, Italy\\
}

\pubyear{2019}
\begin{document}
\label{firstpage}
\pagerange{\pageref{firstpage}--\pageref{lastpage}}
\maketitle

\begin{abstract}
We study the dependency of the concentration on mass and redshift   using three large N-body cosmological hydrodynamic simulations carried out by the Magneticum project.
We constrain the slope of the mass-concentration relation with  an unprecedented  mass range for hydrodynamic simulations \corrd{and find a} negative trend on the mass-concentration plane and a slightly negative redshift  dependency, in agreement with observations and other numerical works.
We also show how \corrd{the concentration correlates} with the  fossil parameter, defined as the  stellar mass ratio between the central galaxy and the most massive satellite, \corrd{in agreement with observations}.
\corrd{We find that haloes with high fossil parameter have systematically higher concentration and investigate the cause in two different ways.
First we study the evolution of haloes} that lives unperturbed for a long period of time, \corrd{where we} find that the  internal region  keeps accreting satellites as  the fossil parameter increases and the scale radius decreases \corrd{(which increases the concentration)}.
We also study  the dependency of the concentration on the virial ratio and the energy term from the surface pressure $E_s$.
\corrd{We conclude} that \corrd{fossil objects have higher concentration because they are dynamically relaxed, with no in-fall/out-fall material and had time to accrete their satellites. }
\end{abstract}

\begin{keywords}
cosmology: dark matter - galaxies: halos - methods: numerical
\end{keywords}

\section{Introduction}


Most density profiles of dark matter haloes  from both simulations and observations can be described  {using} a Navarro Frank and White (NFW) profile (\cite{1996ApJ...462..563N, 1997ApJ...490..493N}, see \cite{2011ASL.....4..204B} for a review).
Such density profile is  {modelled} as a function of the radial distance $r$ as:
$$\rho\left(r\right)=\frac{\rho_0}{\frac{r}{r_s}\left(1+\frac{r}{r_s}\right)^2},$$
where   $r_s$ is a scale radius separating the  { internal and the external regions, and $\rho_0$ is four times the density at $r=r_s$.}

As haloes do not have well defined boundaries, the virial radius $R_{vir}$ is assumed to be the radius at which the density of the halo is the one of a theoretical virialised  spherical overdensity in an expanding universe.
The density threshold is represented as $\Delta_{vir}\rho_{crit}.$
Here $\rho_{crit}$ is the critical density  $\rho_{crit}\equiv3H^2/4\pi G$ and $\Delta_{vir}$ is  a parameter that depends on  cosmology.
For instance, $\Delta_{vir}\approx178$ in an Einstein de Sitter cosmology (see \cite{2015MNRAS.447.1873N} for a review).
More generally, {in the literature, people prefer to make use of radii definitions that are independent of cosmology} and refer to $R_{\Delta}$  as the radius  that includes an over-density of $\Delta\cdot\rho_{crit}.$
In the following analysis, we use  both $\Delta={200}$ and $\Delta={500}$ and the corresponding radii $R_{200}$ and $R_{500}$. 

The concentration $c_{\Delta}$ is defined as $c_{\Delta}\equiv R_{\Delta}/r_s$ and quantifies how wide is the internal region of the cluster, compared to its radius.
\cite{2001MNRAS.321..559B} is a pioneering theoretical work {devoted to} the study of the concentration in a $\Lambda CDM$ universe.
Their toy model based on an isolated spherical over-density, whose scale factor $a$ at the collapse time is $a_c,$  
predicts a concentration $c\propto a/a_c,$ where the proportionality constant is universal for all haloes.

 Various  literature works  make a fit of the concentration as a  power law of the halo mass and redshift.
They mainly found a very low dependency of concentration on redshift and a slow but steady decrease of the concentration with mass \citep[see e.g.][]{2014MNRAS.441.3359D,2015ApJ...806....4M}.
In comparing various works one must first consider carefully how the concentration is computed.
\corr{Some theoretical works  \citep{2012MNRAS.427.1322L, 2012MNRAS.423.3018P} derive the  concentration from the  \corr{circular velocity peak} instead of constraining $r_s$ from a NFW fit of the dark matter density profile.}
\corrd{The concentration derived from the circular velocity peak can have errors up to $1-10\%$ \citep[see][who show how these methods can differ significantly]{2013arXiv1303.6158M}. }
\corr{This  error signals deviations  of the density profile from a purely NFW profile. }

\corr{ The concentration in hydrodynamic simulations is computed by performing a NFW fit on the sole dark matter component of a halo, which itself is influenced by the physics of baryons. 
In fact, stellar and active galactic nuclei (AGN) feedback proved to be able to   transfer momentum to dark matter particles \citep{2016ApJ...820..131E,2018MNRAS.478..906C}.}

\protect\cite{2006ApJ...651..636L} found that introducing non-radiative gas physics  in  numerical simulations increases the concentration,
while \cite{2010MNRAS.405.2161D} showed how the  additional inclusion of AGN feedback decreases the halo concentration \corr{(of  the dark matter density profile)} up to $\approx 15\%$ for haloes with a mass of $\approx10^{11}\Msun.$
\corr{This is in agreement with recent high resolution hydrodynamic simulations, as the  NIHAO simulations \citep{2015MNRAS.454...83W},  where  gas particle masses reach $3\cdot10^3\Msun$.
\cite{2016MNRAS.456.3542T} show how, in the context of  NIHAO simulations,  dark-matter only (DMO) runs produce cuspier  dark-matter profiles.
Additionally, \cite{2016MNRAS.462..663B}   find a flattening of the inner-part of low-mass dark matter density profiles in DMO runs (with respect to NIHAO hydrodynamic simulations).}

Simulations with various dark energy models, as in \cite{2004A&A...416..853D,2013arXiv1302.2364D,2013MNRAS.428.2921D}, showed that the $c-M$ relation  normalisation is sensitive to the  cosmological parameters  and \cite{2008MNRAS.390L..64D}  showed that  the predicted concentrations of dark matter only runs are much lower than {the ones} inferred from X-ray observations of groups and clusters of galaxies.

Additionally,  concentration   inferred from  weak and strong lensing observations can be over estimated due to  intrinsic projection effects \citep{2007A&A...461...25M} or due to the presence of massive background structures \citep{2012ApJ...757...22C}.
When these effects are not correctly  taken into account, concentration  can increase up to  $5-6\%$  and the mass estimation can vary up to $10\%$ \citep{2012MNRAS.426.1558G}.

Most recent high resolution dark matter only simulations showed an upturn trend  in the highest  mass regime of the mass-concentration relation of simulations at very high redshift  \citep[see][]{2009ApJ...707..354Z, 2011ApJ...740..102K,2012MNRAS.423.3018P}.
The cause of such upturn is still unclear.

The mass-concentration relation of various theoretical and observational studies has a scatter   that can span over one order of magnitude.
\cite{2007MNRAS.378...55M} proposed that the scatter is partially due the non spherical symmetry of the initial fluctuations,
while \cite{2007MNRAS.381.1450N} (see Fig. 10 in {their} paper) showed how
this scatter can be partially justified by describing the concentration as a
function of the formation time of the halo.
\corr{The mass accretion history has been found to influence the concentration in
several theoretical works \citep[see e.g.][]{2018arXiv181009473R,2018ApJ...857..118F,2018ApJ...863...37F}.}

Observational studies found that fossil objects (i.e. objects with a dominant central galaxy, compared to its satellites) are the objects with the highest value on concentration \citep[see][]{2016A&A...590L...1P,2015MNRAS.454..161K,2006MNRAS.369.1211K,2012ApJ...755..166H,2011ApJ...729...53H,2017ApJ...834..164B}. 
{This is in agreement with} theoretical studies {on unperturbed haloes in  dark matter only simulations, where    dynamically relaxed haloes have higher concentration than average} \citep{2016MNRAS.457.4340K}.
\nuovo{There are two major hypotheses on the origin of fossil groups: (i) they are
``failed groups'' formed in an environment that lack of massive
satellites (and thus they never had major mergers) or (ii) they are   old systems that  exhausted their
bright satellites through multiple major mergers \citep[see][for more
  details]{2018A&A...618A.172C}}.

Recent works \citep[see e.g.][]{2013ApJ...766...32B}  fit the concentration as a function of the so called "peak height" $\nu,$ where $\nu\left(M, z\right) \equiv \delta_{crit}\left(z\right)/\sigma\left(M, z\right),$  $\delta_{crit} = 1.686$ is the critical density of a collapsing spherical top hat  \citep{1972ApJ...176....1G} and $\sigma\left(M, z\right)$  is the root mean square density of matter fluctuations over a scale $\propto M^{1/3}$ and redshift $z.$ 
This  relation is very useful for theoretical studies (e.g. dependency between concentration and accretion history).
However,  comparisons between theory and observations are usually made by comparing  mass-concentration relations.

Recent observational studies obtain the density profile of the dark matter component inferring the density profile of the baryon component  from X-ray data and remove such component  from the total density profiles obtained with gravitational lensing measurements \citep{2015ApJ...814..120D,2015ApJ...806....4M}.

In this work we analyse the concentration of haloes of the \href{http://www.magneticum.org}{Magneticum}   project {suite} of simulations \citep{2015MNRAS.451.4277D,2016MNRAS.463.1797D}.
The Magneticum project produced  a number of hydrodynamic simulations with different resolutions and  ran  over different volumes including  also   dark matter    runs. 

The plan of this paper is as follows.
\corrd{The selection of haloes is discussed in Section \ref{sec:numsim}.}
In Section \ref{sec:depmass} we fit the concentration as a function of mass and redshift and compare our results with other observational and theoretical works.
In Section \ref{sec:shift}   \corrd{we   fit the concentration as a function of the fossil parameter, and follow the time evolution of fossil objects}.
In Section \ref{sec:virial} we discuss the connection between the concentration and the virial ratio, the energy term from the surface pressure and the fossil parameter.
We summarise our conclusions in Section \ref{sec:concluz}.


\section{Numerical Simulations}
\label{sec:numsim}

\begin{table*}
\caption{Individual setup of the three Magneticum simulations used in this work. The columns contain the name, the box size, the total number of particles, the mass of each dark matter particle, the initial mass of gas particles, the gravitational softening length of  both dark matter and gas  $\epsilon$, and the gravitational softening length of star particles  $\epsilon_\star$ respectively. }
\begin{tabular}{ l r r r  r r r r}
Simulation\\
  Name & Size      & n. part &  $m_{dm}$     & $m_{gas}$   &  $\epsilon$ 	   & $\epsilon_{\star}$  \\
       & $[Mpc/h]$ &         &  $[\Msun/h]$  & $[\Msun/h]$    &  $[kpc/h]$	        &  $[kpc/h]$ \\
  \hline
  Box4/uhr & 48 & $2\cdot576^3$ &   $3.6\cdot10^{7}$	& $7.3\cdot10^{6}$	& 1.4	 	& 0.7 \\
 Box2b/hr & 640 & $2\cdot2880^3$ &  $6.9 \cdot10^{8}$ &	$1.4\cdot10^{8}$ &	3.75 	& 2\\
 Box0/mr & 2688 & $2\cdot4536^3$ &  $1.3 \cdot10^{10}$ &	$2.6\cdot10^{9}$ &	10   &	5 \\
  
 \end{tabular}
\label{table:sims}
\end{table*}

\begin{table*}
\caption{{Number of haloes   in each snapshot, that  have $M_{200}$ higher than minimum mass for resolved haloes (corresponding to at least $10^4$ particles).  } }
 \begin{tabular}{l l l l l l l l}
  \hline
        &      &        redshift                   &  $0$  & $ 0.5$  & $1$  & $1.5$  & $2$  \\
\hline
    Simulation    &   Min      $M_{200}$   & Max $M_{200}$  $(z=0)$               & n. haloes   \\
               &     $[\Msun/h]$              &    $[\Msun/h]$ &      \\

    Box4/uhr & $1.3\cdot10^{11}$ & $1.3\cdot10^{14}$ & 1845  & 1775  & 1934 & 1839  & 1782 \\
   Box2b/hr &  $4\cdot10^{12}$   & $1.8\cdot10^{15}$ & 156110  & 146339  & 99669   & 63542   & 48925  \\
   Box0/mr &   $8\cdot10^{13}$  & $3.8\cdot10^{15}$&  329648 & 140560 &  21274 & 7792 & 1112 \\
  \hline 
  \end{tabular}
\label{table:boxdata}
\end{table*}

The Magneticum simulations \citep[\href{http://www.magneticum.org}{www.magneticum.org},][]{2013MNRAS.428.1395B,
2014MNRAS.440.2610S, 2015MNRAS.448.1504S,
2016MNRAS.463.1797D, 2015MNRAS.451.4277D, 2015ApJ...812...29T,
2016MNRAS.458.1013S, 2016MNRAS.456.2361B, 2017MNRAS.464.3742R} is a set of simulations that follow the evolution  of overall up to $2\cdot10^{11}$ particles of dark matter, gas, stars and black holes  on cosmological volumes.
The simulations were performed with an extended version of the N$-$body/SPH code Gadget3 
which itself is the successor of the code P-Gadget2 \citep{2005Natur.435..629S,2005MNRAS.364.1105S}.
Gadget3 uses an improved Smoothed Particle Hydrodynamics (SPH) solver for the hydrodynamics evolution of gas particles presented in \cite{2016MNRAS.455.2110B}.
\cite{2005MNRAS.361..776S}   describe the  treatment of radiative cooling, heating, ultraviolet (UV) back-ground,  star formation and stellar feedback processes. 
Cooling follows  11 chemical  elements ($H, He,
C, N, O, Ne, Mg, Si, S, Ca, Fe$)  using the publicly available CLOUDY photo-ionisation
code \citep{1998PASP..110..761F} while \cite{2010MNRAS.401.1670F,2014MNRAS.442.2304H} {describe prescriptions for black hole growth and for feedback from AGNs} .
 
Galaxy haloes are identified using a friend-of-friend (FoF) algorithm and sub-haloes are identified using  a version of SUBFIND
\citep{2001MNRAS.328..726S}, adapted by  \cite{2009MNRAS.399..497D} to include the baryon component.  
\corr{This SUBFIND version additionally computes  the values of $M_{200}$ and $M_{500}$ that are used in this work.} 

{The simulations assume a cosmological model in agreement with the WMAP7 results \citep{2011ApJS..192...18K}, with total matter density parameter $\Omega_{0,m}=0.272,$ a baryonic fraction of  $16.8\%,$ Hubble constant $H_0=70.4\ km/s/Mpc,$ index of the primordial power spectrum $n=0.963$ and  a normalisation of the power spectrum corresponding to  $\sigma_8=0.809.$}

 In particular, we use three of the Magneticum simulations presented in Table  \ref{table:sims}. 
  We use Box0/mr to follow the most massive haloes, Box2b/hr  to follow  haloes  within an intermediate mass range   and Box4/uhr to follow haloes with masses in the  galaxy range. 
  
\corr{The detailed description of baryon physics in Magneticum simulations
is capable of matching several observed properties of galaxies and their haloes.
For instance: the specific angular momentum
 for different morphologies \citep{2015ApJ...812...29T,2016ilgp.confE..41T}; the  mass-size relation
\citep{2016ilgp.confE..43R,2017MNRAS.464.3742R,2019MNRAS.484..869V};   the dark matter fraction \citep[see Figure 3 in][]{2017MNRAS.464.3742R}; 
the  baryon conversion efficiency  \citep[see Figure 10 in][]{2015MNRAS.448.1504S};   kinematical observations of early-type galaxies  \citep{2018MNRAS.480.4636S};  the  inner slope of the total matter density profile  \cite[see Figure 7 in][]{2018MNRAS.476.4543B}, the ellipticity and velocity over
velocity dispersion ratio  \citep{2019MNRAS.484..869V}.}

From each simulation we selected snapshots  nearest  to redshifts $z\approx0, 0.5, 1, 1.5$ and $2$.
In each snapshot we chose only haloes with a number of dark matter particles greater 
than $10^4.$ 
\corr{Numerical studies show how  $\approx10^4$ particles are enough for a convergence of the NFW fit \citep{1998ApJ...499L...5M}.
We subsequently apply a cut in the critical mass so that all objects within this cut are well resolved.}

Table \ref{table:boxdata} lists the number  of  selected haloes, for each simulation and redshift, that match this mass-cut criterion.


\section{The dependency of concentration on mass and redshift }
\label{sec:depmass}

For all selected Magneticum  haloes  in Table  \ref{table:boxdata}, we fit the concentration as a function of mass, using  the following functional form: 
\begin{equation} \label{eq:fit_c}c_{200}=A\cdot\left(\frac{M_{200}}{10^{13}\Msun}\right)^B.\end{equation}

\begin{figure*}
\centering
\includegraphics[width=0.47\textwidth]{\mypath{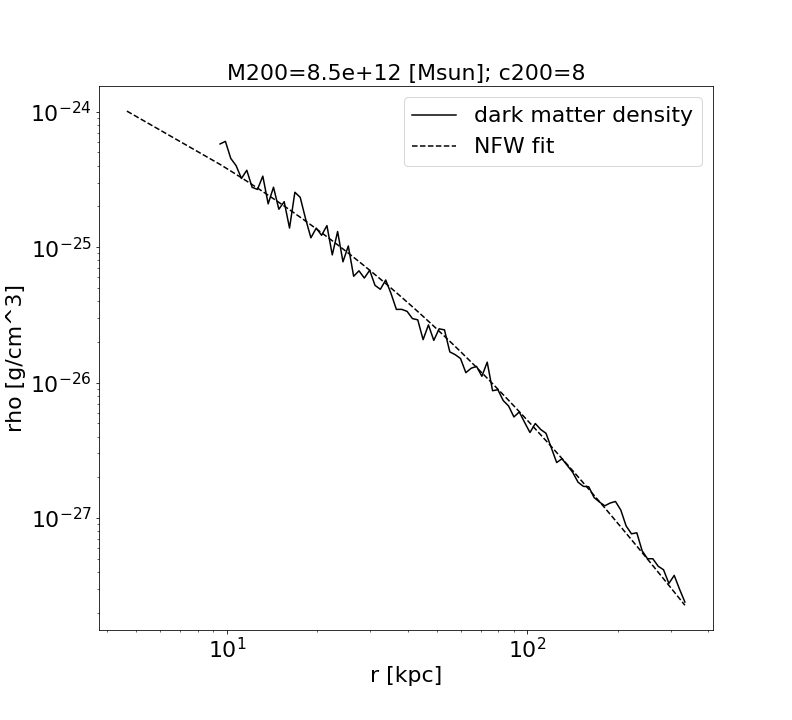}}
\includegraphics[width=0.47\textwidth]{\mypath{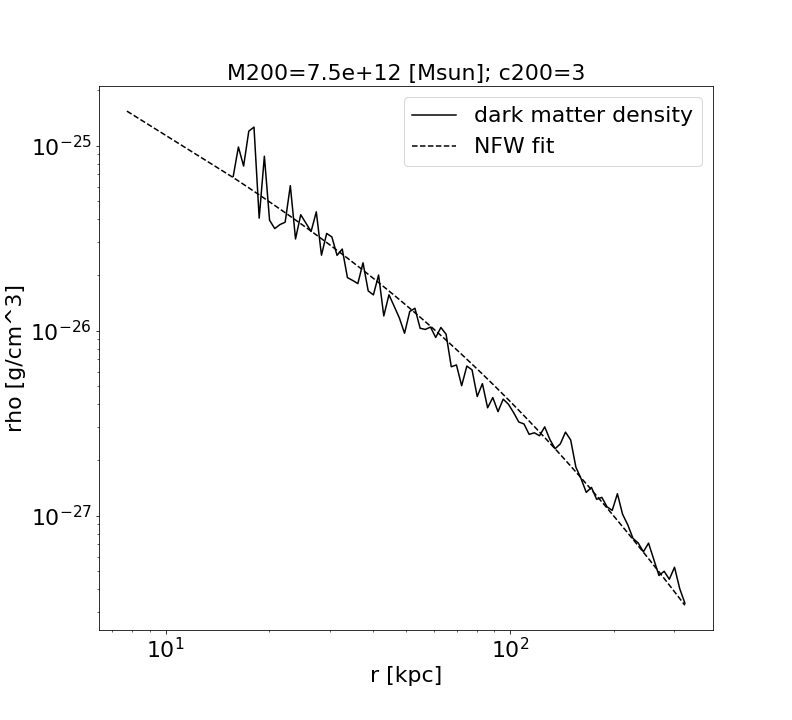}}
\caption{\corr{Dark matter density distribution and NFW fit profile of two haloes, one with  a high concentration (left panel) and one with a low concentration (right panel)}}\label{fig:seefit}
\end{figure*}

\corr{The  scale radius $r_s$ is order of magnitudes above the resolution limit of the simulation (for instance, the softening lengths $\epsilon$ and $\epsilon_\star$ in Table \ref{table:sims}), making it a well resolved value.  
The NFW fit is performed over $50$ logarithmic bins of the  dark matter density, up to $R_{200}.$  The first bin runs from the centre of the halo to the minimum distance that contains $100$ particles.
Figure \ref{fig:seefit} shows the dark matter density profile and the corresponding NFW fit for a low concentrated and a high concentrated halo.}

We performed the fit for various redshift bins $z=0,0.5,1,1.5,2$ and over the whole range $z=0-2$. 
The fit was performed  using the average  concentration computed in  $20$ logarithmic mass bins that span the whole mass range.
The pivot mass $10^{13}\Msun$ is the median mass of all selected haloes.

When we extract all haloes in a mass range over different snapshots  from a simulation, it happens that  most haloes at high redshift will be re-selected at lower redshift.
We argue that this does not introduce a bias in the selection: in fact, the time between the two snapshots is longer than the dynamical time of the halo, ensuring that there is no correlation between the dynamical states of the two objects after such a long period of time.

\corr{We then fit the concentration as  a function of both mass and redshift, with the following functional form:}
\begin{equation} \label{eq:fit_c_mz}c_{200}= A \cdot \left(\frac{M_{200}}{  10^{13} \Msun}\right)^B \left(\frac{1.47}{1+z}\right)^C.\end{equation}

\begin{table*}
\caption{Fit parameters  of $c_{200}(M_{200})$ as a power law of the halo mass as in Equation \ref{eq:fit_c} for each redshift bin.}
 \begin{tabular}{l l l}
  redshift & A &  B   \\ 
  \hline
$ z = 0$ & $6.25 \pm 0.07$  &  $-0.121 \pm 0.004$\\
$ z = 0.5$ & $5.79 \pm 0.07$  &  $-0.122 \pm 0.004$\\
$ z = 1$ & $5.26 \pm 0.08$  &  $-0.123 \pm 0.007$\\
$ z = 1.5$ & $5.36 \pm 0.07$  &  $-0.117 \pm 0.006$\\
$ z = 2$ & $5.37 \pm 0.07$  &  $-0.097 \pm 0.006$\\
$ z = 0-2$ & $5.74 \pm 0.07$  &  $-0.104 \pm 0.004$\\
\hline 
  \end{tabular}
\label{table:fitzbins}
\end{table*}

\begin{figure*}
\centering
\includegraphics[width=0.65\textwidth]{\mypath{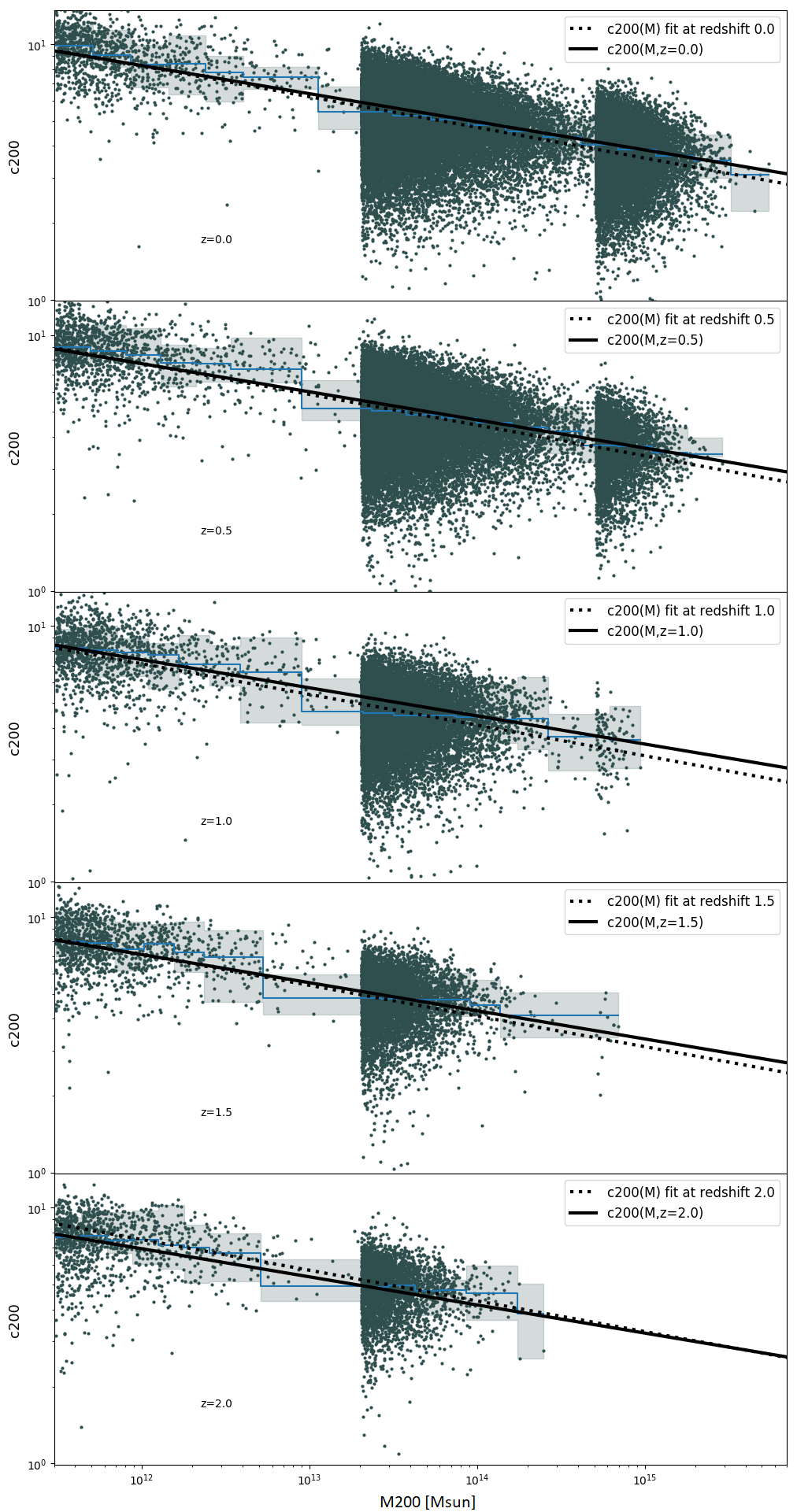}}
\caption{Mass-concentration relation for the well resolved haloes in the three Magneticum simulations Box4/uhr, Box2b/hr and Box0/mr  (dark points). Each  simulation covers  three different mass ranges, respectively $M_{200}>3\cdot10^{11}\Msun, M_{200}>2\cdot10^{13}\Msun$ and $M_{200}>5\cdot10^{14}\Msun$  .  In each panel we show haloes of a  different redshift bin, the median  of the concentration(blue curve), the locus containing $50\%$ of points (shaded area), the fit obtained with a $c_{200}(M_{200})$ fit as in Equation \ref{eq:fit_c}  and $c_{200}(M_{200},z)$ as in Equation \ref{eq:fit_c_mz} (dotted and solid  lines, respectively).}\label{fig:ciao}
\end{figure*}

Table \ref{table:fitzbins} shows the   fit parameters and their errors that are given by the cross-correlation matrix.
The  concentration at $10^{14}\Msun$ evolves very weakly with redshift.
In order to confirm this, for all selected haloes presented  in Table \ref{table:boxdata}, we also  performed a fit of the halo concentration as a power law of mass and redshift using the relation

The fit was made on the average concentration of the haloes binned by the $5$ redshift bins  on the same mass bins as before and  for the redshift dependency we use the median redshift value of $1.47$ as pivot.
The fit,  performed  over all objects gives:

\begin{equation}
\begin{split}
 A  = &6.02 \pm 0.04\\
B  =   &-0.12 \pm 0.01\\
C    =  & 0.16 \pm 0.01\\
\end{split}
\end{equation}

We can see that the redshift dependency, {represented by the parameter C},  is low although  it  differs from zero.

Figure \ref{fig:ciao} shows the mass-concentration plane of Magneticum haloes, where different panels display data at different  redshifts.
Over-plotted are the fit relations for $c_{200}\propto M_{200}^B$ and  $c_{200}\propto M_{200}^B\cdot (1+z)^{-C}.$

\begin{table*}
\caption{Mass ranges and fit parameters  of the mass-concentration  relation in literature. The value of $c_{200}(10^{14}\Msun)$ is extrapolated at  $z=0$  when the relative error in the fit parameter is smaller than few percents (when provided). { Concentration in \protect\cite{2001MNRAS.321..559B}  has been converted from  $c_{vir}$ to $c_{200}.$} }\label{tab:mcznoca}
\begin{tabular}{ l r c l r r l }
  authors &  \multicolumn{3}{c}{  mass range $[\Msun]$}  & slope & $c_{200}(10^{14}\Msun)$   & comments \\
         &      &&     &  &    & \\\hline

  \cite{2001MNRAS.321..559B} &  $10^{11}$ &$-$& $10^{14} h^{-1}$         &  $\approx -0.3$ &  $4.1$ & N-body  \\
  {\citet{2005A&A...429..791P}} &  $10^{14}$ &$-$& $10^{15}$           &  $N/A$     & $4-6$ &  X-ray from XMM-Newton \\
  \cite{2007MNRAS.381.1450N} &  $10^{12} $&$-$&$ 10^{15}  h^{-1}$          & $-0.1$ &   $4.8$ & N-body from Millennium \\
  \cite{2008JCAP...08..006M} &  $10^{12} $&$-$&$ 10^{15}  h^{-1}$          & $-0.13\pm0.07$    & $4.8$ &  weak lensing via SDSS\\
  \cite{2013ApJ...766...32B} &  $\sim3\cdot10^{12}$&$-$&$  10^{15}  h^{-1}$ &  $-0.08$ & $4.7$ & N-body \\
 {\cite{2013A&A...557A.131M}} &  $ 10^{11}$&$-$&$  10^{12}$ &  $N/A$ & $N/A$ & Subset of DiskMass survey\\
 \cite{2014MNRAS.441.3359D}   &  $10^{12.5} $&$-$&$  10^{14.5} h^{-1}$   & $-0.905$  & $5.2$ & N-body \\
  \cite{2014ApJ...797...34M} &  $6\cdot10^{14}$&$-$&$ 10^{15}  h^{-1}$ & $-0.058$ & $N/A$ & CLASH mock observations \\
  \cite{2014MNRAS.441..378L} &  $10^{12}$&$-$&$ 10^{15}  h^{-1}$            & $-0.1$ & $5.5$ & N-body from Millennium  \\
  \cite{2014ApJ...784L..25C}  &  $3\cdot10^{13}$&$-$&$2\cdot10^{14}  h^{-1}$  & $0.09$ & $5.4$ & lensing from CFHTLenS  \\
  \cite{2015MNRAS.452.1217C} &   \multicolumn{3}{c}{$N/A$}  & $-0.08$ & $3.8$ &  semi-analytical model\\
  \cite{2015ApJ...806....4M} &$5\cdot 10^{14}$&$-$&$2\cdot10^{15} $ &  $-0.32\pm0.18$ & $N/A$ & lensing+X rays on CLASH data\\

  \cite{2016MNRAS.462..681M}  & $5\cdot10^{14}$&$-$&$2\cdot 10^{15}  $  & $-0.15$ & $N/A$ &  lensing and X-ray from Chandra and ROSAT \\
 \cite{2016MNRAS.455..892G} &   \multicolumn{3}{c}{$\sim10^{15} $} & $-0.16$ & $N/A$ & comprehensive study on lensing data\\ 
\cite{2016MNRAS.457.4340K} &   $10^{11}$&$-$&$10^{15} h^{-1}$ & $-0.12$ & $4.1$ & N-body from MultiDark   \\
\cite{2017ApJ...840..104S}  & $5\cdot10^{12}$&$-$&$2\cdot10^{14} $  & $-0.13$ & $3.3$ & weak lensing on SDSS/BOSS \\
{\cite{2017A&A...607A..81B}} &   $10^{14}$&$-$&$2\cdot10^{15}  $   &$-0.11\pm0.1$     & $4.6$   & dynamics of OmegaWINGS clusters\\
\cite{2018MNRAS.477.2804S} & $5\cdot10^{14}$&$-$&$2\cdot10^{15} $ & $-0.14$ & $5.6$ & Omega$500$ hydrodynamic simulations \\
  This work                & $10^{11}$&$-$&$10^{15}$  &   $-0.1$ & $4.5$ & Hydro  N-body from Magneticum \\
  \hline
 \end{tabular}
\end{table*}

\begin{figure*}
\includegraphics[width=1.\textwidth]{\mypath{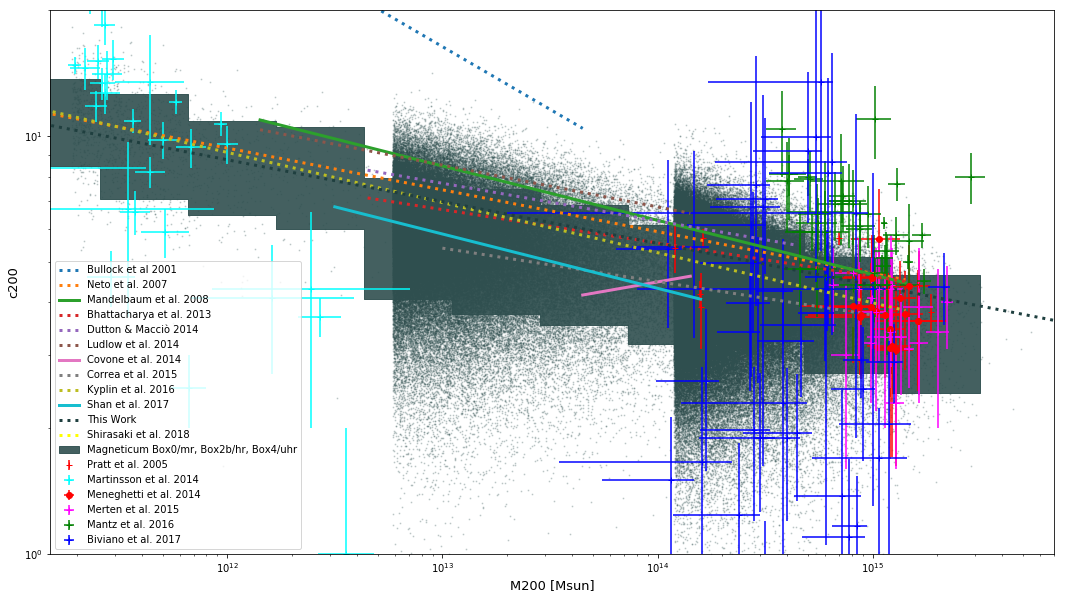}} 
\caption{The mass-concentration relation. Dark green points are haloes from the Magneticum simulations (see Table \ref{table:sims})  at $z=0$, dark shaded area contains $95\%$ of points within the median. Black dashed line is the $c_{200}(M_{200})$ median for Magneticum data points. Dashed lines are predictions from simulations and solid lines are fit from observed concentrations, { both at $z=0$}. Error bars are from observations from dynamical mass analyses, with no redshift corrections. All mass conversions are made assuming $h=0.704.$ } \label{fig:mcznoca}
\end{figure*}

Table   \ref{tab:mcznoca} reports a review of the slope values of the mass-concentration plane found on both theoretical and observational works.
Figure \ref{fig:mcznoca} shows  a plot of the same data.
When the slope of the mass-concentration  relation  had an uncertainty smaller than few percents, we extrapolated the value of the concentration at the mass of  $10^{14}\Msun$   using $h=0.704.$

\cite{2001MNRAS.321..559B} present one of the first analytical and numerical work on concentration in  simulations.
 They predicted the concentration within the virial radius, that in this work has been converted to a concentration over $R_{200}.$
 Although their simulations were performed with a relatively low resolution, their concentration extrapolated at $10^{14}\Msun$ is within the scatter of present days studies.
\cite{2007MNRAS.381.1450N} employ the first very large dark matter-only N-body cosmological simulation, the Millennium simulation,   see \cite{2005MNRAS.364.1105S} where they constrain  the mass-concentration dependency accurately over several orders of magnitudes in mass for dark matter only runs.

 \cite{2005A&A...429..791P} use X-ray data from XMM-Newton,  \cite{2008JCAP...08..006M,2017ApJ...840..104S} use lensing from SDSS images, while \cite{2014ApJ...784L..25C, 2016MNRAS.462..681M, 2016MNRAS.455..892G,2014ApJ...784L..25C, 2016ApJ...821..116U} combine both lensing and X-ray reconstruction techniques to find the concentration of the dark matter component of haloes.
Observations with X-ray data  have usually  high uncertainties and need to make assumptions on the dependency between the baryon   and the dark matter profiles, producing data with large uncertainties.
The low mass regime of the plot shows observations of galaxies from the DiskMass survey from \cite{2013A&A...557A.131M}.
Points from the  DiskMass survey  cover a very large range of concentration values for low massive haloes,   in contrast with    simulations.
\cite{2015MNRAS.452.1217C} adopted a semi-analytical model (SAM) that predicts concentration over 5 orders of magnitude.
\cite{2016MNRAS.455..892G} stack all observational mass-concentration data found in literature and made a single fit from it.
\cite{2016MNRAS.457.4340K} show  the results of the MultiDark N-body simulation and produce a lower concentration than Magneticum haloes.
\cite{2014ApJ...797...34M} present  a numerical work called   MUSIC of CLASH where a number of simulated haloes have  been chosen to make mock observations for CLASH. 
{\cite{2016MNRAS.462..681M} present results from observations of relaxed haloes.
These haloes have a higher concentration in agreement with theoretical studies.}
 The high mass regime of the plot shows results from observations from WINGS  \citep{2017A&A...607A..81B} and from CLASH \citep{2015ApJ...806....4M}.
It must be taken into account that the galaxies from the DiskMass survey are a restricted sub-sample of a very large initial sample.
Those galaxies have been chosen so that it is possible to compute the concentration.
This may have introduced a significant  bias in the concentration estimate.
\cite{2015ApJ...806....4M,2017A&A...607A..81B,2005A&A...429..791P,2013A&A...557A.131M} {compute halo properties using dynamical analyses which have larger uncertainties.}
\nuovo{The Omega500 simulations \citep[see e.g.][]{2018MNRAS.477.2804S} are    hydrodynamic simulation that include radiative cooling, star formation
  and  AGN feedback}.

\corr{Magneticum low-mass haloes have comparatively  lower concentration of the dark matter profile than   dark matter only simulations.}

\section{Concentration and fossil parameter}
\label{sec:shift}

The previous section showed how the concentration can span over an order of magnitude on both observational and theoretical works.
In this section  we show how the scatter is partially related to ``how much'' a halo is fossil.
We first define a fossilness parameter and then study the evolution over time of both the fossilness and the concentration in some special  objects.

\cite{2016A&A...590L...1P,2015MNRAS.454..161K,2006MNRAS.369.1211K,2012ApJ...755..166H,2011ApJ...729...53H,2017ApJ...834..164B} show how fossil objects have a higher concentration than the average.

More generally, simulations  found that dynamically relaxed haloes have a higher concentration
\citep[see e.g.][]{2016MNRAS.457.4340K}.

A fossil object has been defined by \cite{2010ApJ...708.1376V} as having a difference in magnitude in the $R$ band $\Delta m_R \geq 1.7$ between the most luminous object and the second most luminous object within a distance of $\frac{1}{2}R_{200}$ from the centre.

In our theoretical work we adapt the definition of the fossil parameter by quantifying it as the stellar mass ratio between central galaxy and most massive satellite: 
\begin{equation} \label{eq:fossilness}
\rm{fossilness}=\frac{M_{\star,\rm{central}}}{\rm{max}\left\{M_{\star,\rm{satellite}}\right\}}.
\end{equation}

We also extended the search of all satellites to $R_{200}$ (instead of $\frac{1}{2}R_{200}$ proposed by \cite{2010ApJ...708.1376V}) because we consider objects outside $R_{200}$    not to contribute to the dynamical state.

We convert  the  observed magnitude difference to a fossil parameter by assuming a constant ratio between galaxy masses and luminosities, \begin{equation} \label{eq:cfossilness}\rm{fossilness}=10^{\Delta m_R/2.5}.\end{equation}

This implies that the $\Delta m_R \geq 1.7$  threshold defined in \cite{2010ApJ...708.1376V} corresponds to a fossilness of  

\begin{equation} \label{eq:hifossilness}\rm fossilness \gtrsim4.5.\end{equation}

\subsection{Concentration as a function of the fossil parameter}

\begin{figure*}
\centering
\includegraphics[width=\textwidth]{\mypath{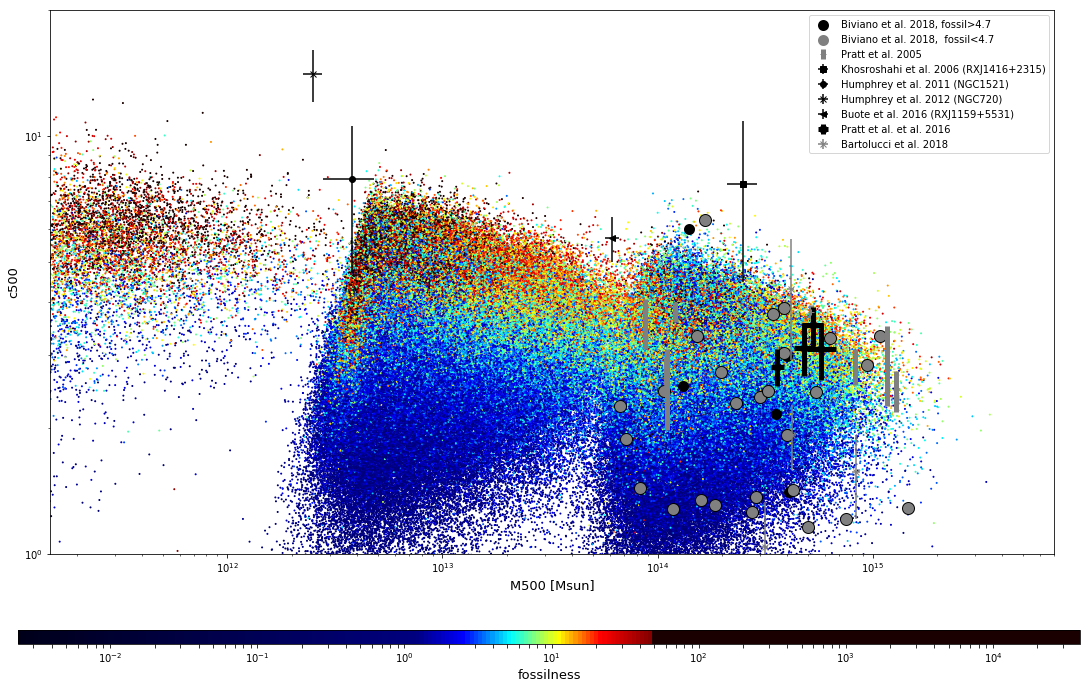}}
\caption{Mass-concentration plane. Mass and radius are computed using $\Delta_{500}$. Points are from the Magneticum data and they are colour coded by fossil parameter (defined as the ratio between stellar masses of the central galaxy and the most massive satellite).  {The colour saturates to black for the $10\%$ outliers in concentration.} Fossil objects from \protect\cite{2006MNRAS.369.1211K, 2011ApJ...729...53H,2012ApJ...755..166H,2016A&A...590L...1P,2017ApJ...834..164B} are coloured in black, haloes from \protect\cite{2005A&A...429..791P,2018A&A...617A..64B,2017A&A...607A..81B} are coloured in grey. Data from \protect\cite{2017A&A...607A..81B} is divided between high and low fossilness according to Equation \ref{eq:hifossilness}. }
\label{fig:mcfossil}
\end{figure*}

Figure \ref{fig:mcfossil} shows the Magneticum haloes concentration as a function of halo mass, colour coded by fossilness.
 We also show observational data of fossil groups   taken from \cite{2006MNRAS.369.1211K, 2011ApJ...729...53H,2012ApJ...755..166H,2016A&A...590L...1P,2017ApJ...834..164B} and haloes from \cite{2005A&A...429..791P, 2017A&A...607A..81B,2018A&A...617A..64B}.
 Since most  observational data were provided in terms of $R_{500}$ and $c_{500},$ in this plot we show mass and concentration computed using $\Delta=500$ for all data points.
 Haloes from \cite{2017A&A...607A..81B} are colour coded by fossilness by converting the difference in magnitude to ratio of luminosities.

\begin{figure}
\centering
\includegraphics[width=.45\textwidth]{\mypath{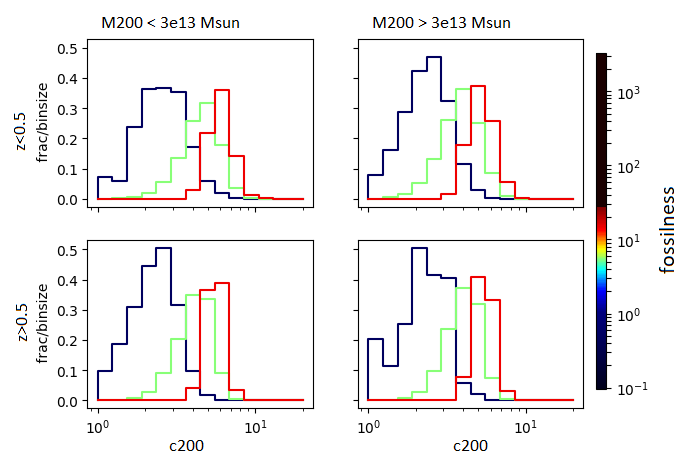}}
\caption{Distribution of concentration for various fossilness values.  Left panels contain low mass haloes ($M_{200}<3\cdot10^{13}\Msun$) and right panels contain high mass haloes  ($M_{200}>3\cdot10^{13}\Msun$) , while top row  refers to low redshift haloes  ($z\leq0.5$)  and  bottom row refers to high redshift haloes   ($z>0.5$).
}\label{fig:mcMsatMcdBins}
\end{figure}

Figure \ref{fig:mcMsatMcdBins}  shows   the concentration distribution for various mass, redshift and colour coded by fossilness bins.
We can see that at each mass and redshift bin, the concentration increases with  the fossil parameter, while the spread decreases as the fossil parameter increases.

\begin{figure*}
\centering
\includegraphics[width=.95\textwidth]{\mypath{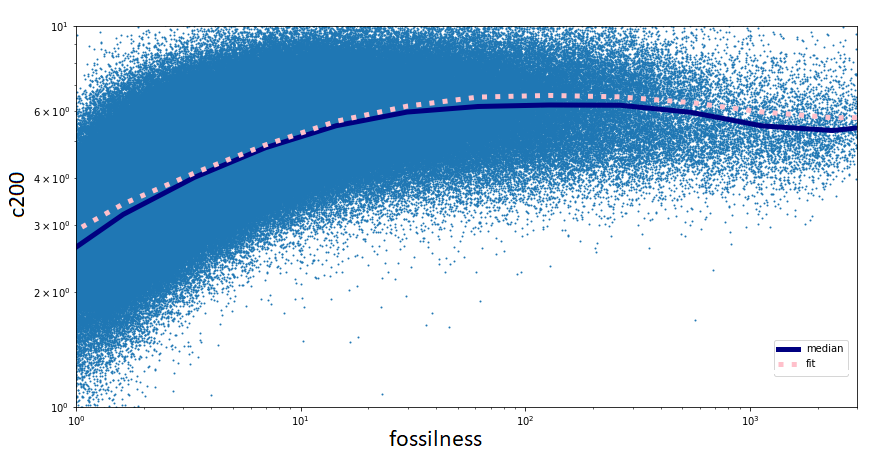}}
\caption{Concentration vs. fossilness for Magneticum data. Over plotted are the median, the concentration depending   a double power law of the fossilness.}\label{fig:checkfitall}
\end{figure*}

There is a change in slope for very high value of the fossilness parameter  so we modelled the dependence of concentration with     slopes (see Figure  \ref{fig:checkfitall}, with also the fit results):

\begin{equation} 
\begin{split}
c_{200}=& A \cdot \left(\frac{M_{200}}{ 10^{13} \Msun}\right)^B\left(\frac{1.47}{1+z}\right)^C   \cdot \\
&\cdot\left(\frac{\rm{fossilness}}{F_0}\right)^{D } \left(1+\frac{\rm fossilness}{F_0}\right)^{E-D}.
\end{split}
\label{eq:fossilnessola}
\end{equation}

The fit was performed with the binning technique as for the previous fits.
Additionally, the fossil parameter was binned over $20$ logarithmic bins of $fossilness>1.$
In this case, the exponent  $E$ maps the asymptotic exponent of $c_{200}$ for high values of fossil parameters, while $D$ is the exponent for   low values of the fossil parameter.
The value of $F_0$ in the fit should should indicate where the two regimes of the fossilness slope starts to change.

\begin{table*}
\caption{\corrd{Fit parameters  of $c_{200}(M_{200})$ as a function of mass, redshift and fossil parameter as in Equation \ref{eq:fossilnessola}.}}
 \begin{tabular}{r l }
\hline
Fit parameter & Value\\
\hline
$A$ & $7.5  \pm 0.1 $\\
$B$ & $-0.1 \pm 0.1 $\\
$C$ & $0.13 \pm 0.01 $\\
$D$ & $0.40 \pm 0.03 $\\
$E$ & $-0.015 \pm 0.003 $\\
$F_0$ & $4.8 \pm 0.7 $\\
\hline
\end{tabular}
\label{table:fitfoss}
\end{table*}

Table \ref{table:fitfoss} show the fit results.
\corrd{There it is possible to see the positive correlation between concentration and fossilness (parameters $D$ and $D-E$ are positive).}
Figure   \ref{fig:checkfitall}  shows the  fitting relation as well as   the data for single haloes and their median.
For higher values it is necessary to use a double slope relation.

\subsection{Concentration evolution in time}

\begin{figure*}
\centering 
\includegraphics[width=.49\textwidth]{\mypath{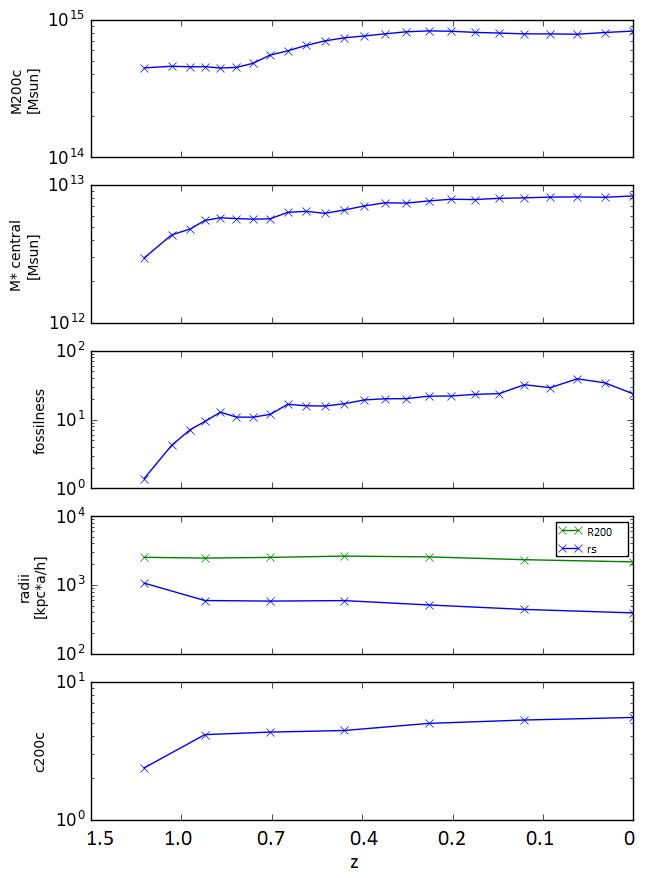}}
\includegraphics[width=.49\textwidth]{\mypath{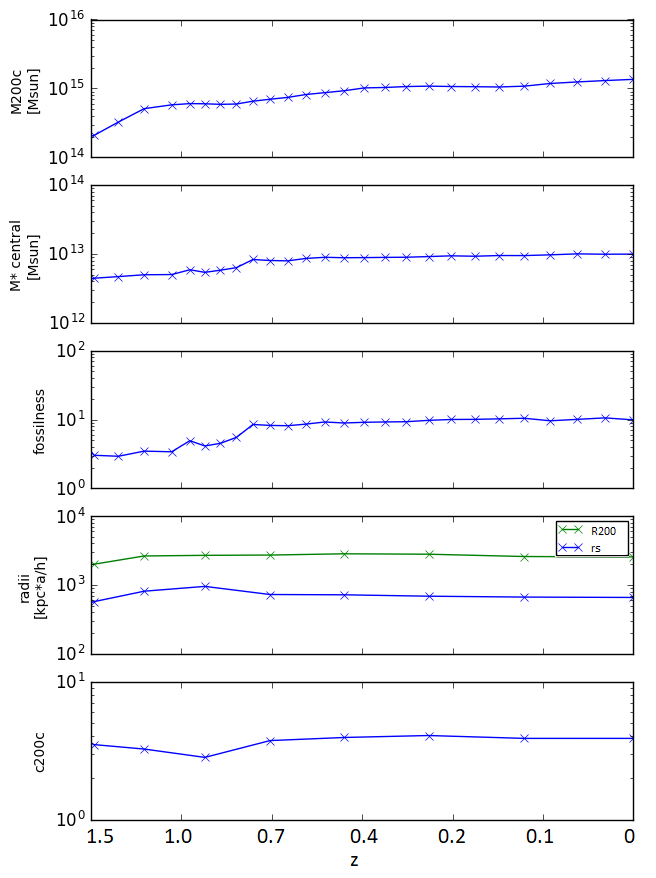}}
\caption{The evolution over time of two haloes (left and right panels) from Box0/mr: $M_{200},$ stellar mass of central galaxy,  fossil parameter, $R_{200}$ (in green) and $r_s$ in (blue) and concentration from to to bottom. Both objects have been selected because they had an increasing fossil parameter. As long as their central galaxy accretes satellites and keep accreting mass, the scale radius decreases and in turn, decreases the  concentration to decrease, thus the relationship between concentration and fossilness.}
\label{fig:c_grow}
\end{figure*}

\begin{figure*}
\centering
\includegraphics[width=.49\textwidth]{\mypath{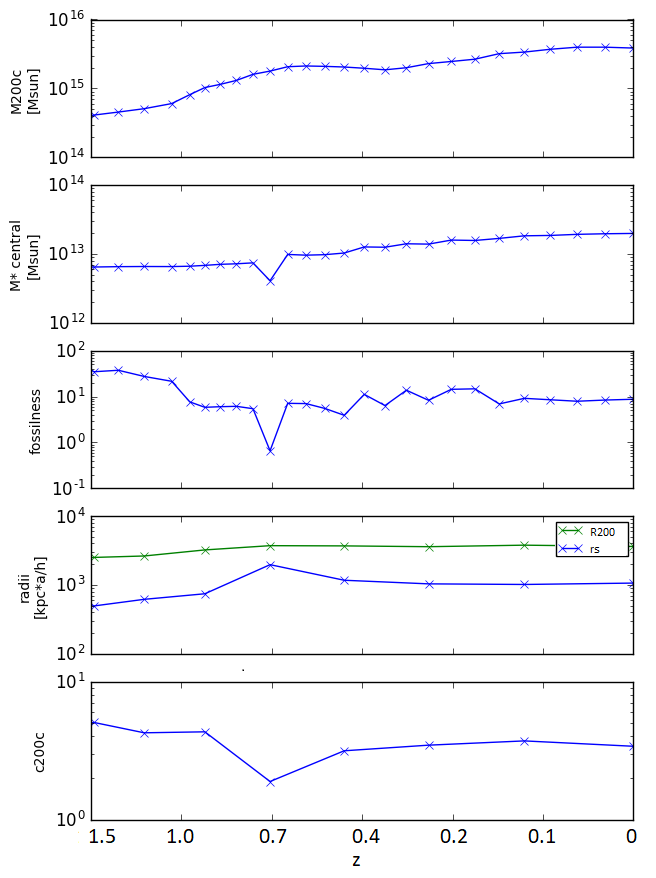}}
\includegraphics[width=.49\textwidth]{\mypath{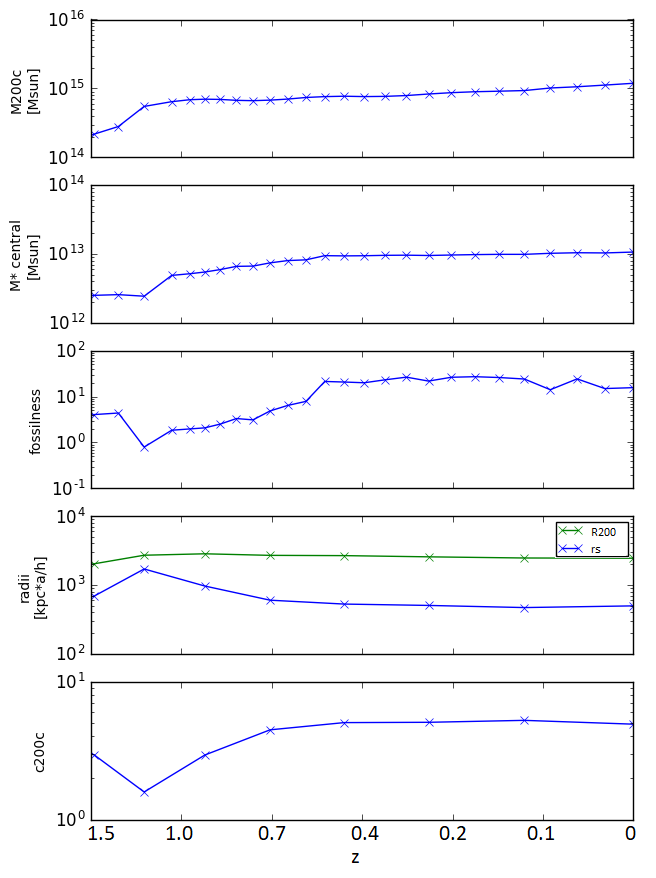}}
\caption{As for Figure \ref{fig:c_grow} but for objects that have a single major merger in their history. }
\label{fig:c_event}
\end{figure*}

In order to understand  what brought fossil objects such a high concentration, we  followed the evolution of concentration and fossilness for a number of objects in  the simulation Box/0mr.
We present here two of the few most massive objects where fossil parameter increased  from $z=1.5$  to $z=0$.
They have more than $10^{5}$ particles and a final mass $M_{200}\approx10^{15}\Msun.$
Figure \ref{fig:c_grow}  shows the evolution of halo mass, the stellar mass of central galaxy, scale radius, halo radius, fossilness and concentration of {  these haloes}.
In these examples it is very easy to see that as long as their central galaxy accretes satellites  and keeps accreting mass, the scale radius decreases and makes their concentration higher and higher.

Additionally, in Figure \ref{fig:c_event} we show  the evolution  of two haloes that happen to have only one major merger in their history.
{When a merger happens then the fossil parameter drops because  new massive satellites enter the system and the fossilness value decreases (see Eq. \ref{eq:fossilness}). 
As already expected from previous theoretical studies \citep{2007MNRAS.381.1450N} we can see that the concentration goes down.}

\cite{2007MNRAS.381.1450N} showed how the scatter in concentration can be partially described by the formation time,
in this subsection we showed how a shift in concentration caused by a slow and steady increase of the concentration (led by a decrease of $r_s$)  brings future fossil groups in the top region of the mass-concentration plane.

\section{Virial ratio and concentration}

In this section we study how the virial ratio of Magneticum haloes depend on the concentration and fossilness.

The moment of inertia $I$ of a collisionless fluid under a force given by its gravitational potential $\Phi$, obeys the time evolution equation: 

$$ \frac{1}{2} \frac{d^2I}{dt} = 2K + W  - E_s,$$

where the kinetic energy $K$ includes the internal energy of gas, $W$ is the total potential energy of the system and
$E_s$ is  the energy from the surface pressure $P$ at the halo boundary: 
$$ E_s = \int_S P(\vec r) r\cdot d\vec S.$$
The pressure takes into account the pressure from the gas component.

A system at the equilibrium is supposed to have the so called {virial ratio $\eta=1,$ where} 

$$\eta \equiv -\frac{2K-E_s}{W}.$$

For  more details on how to compute these quantities and integrals see \cite{1961hhs..book.....C,2008gady.book.....B,2017MNRAS.464.2502C}.

\label{sec:virial}
\begin{figure*}
\centering
\includegraphics[width=.49\textwidth]{\mypath{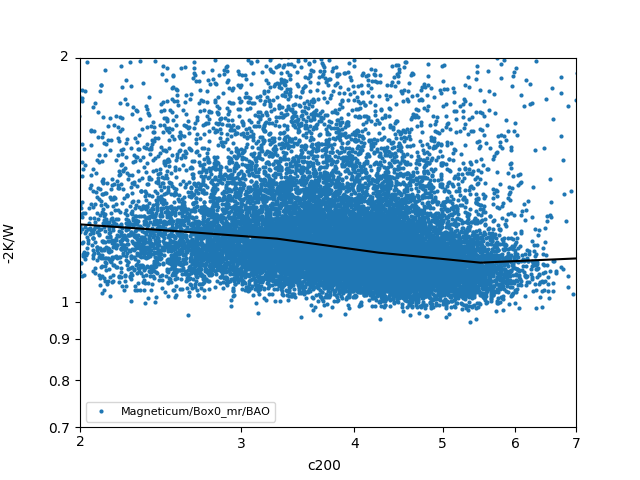}}
\includegraphics[width=.49\textwidth]{\mypath{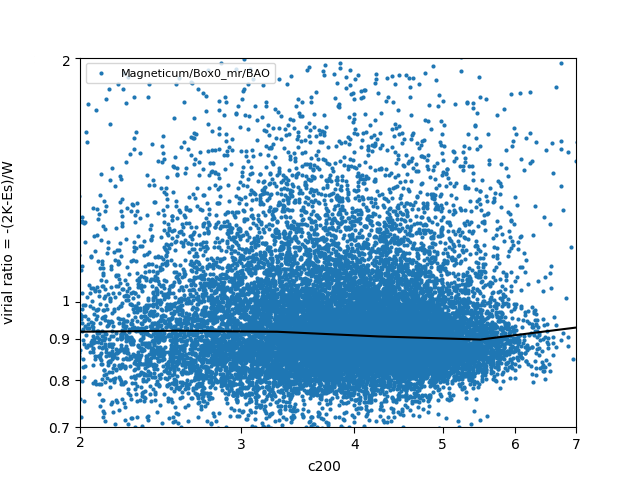}}
\caption{Virial ratio without (left panel) and with (right panel) the correction from the pressure term, as a function of the concentration for the simulation Magneticum/Box0/mr.}\label{fig:virial}
\end{figure*}

\begin{figure}
\centering
\includegraphics[width=.55\textwidth]{\mypath{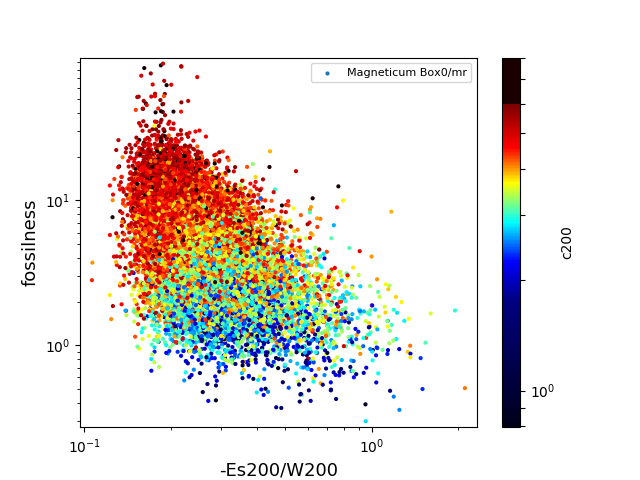}}
\caption{Fossilness versus virial ratio for Magneticum/Box0/bao run, colour coded by concentration (colour bar on the right). The colour saturates to black for the $10\%$ outliers in concentration.}\label{fig:virialfossil}
\end{figure}

Figure \ref{fig:virial} (left panel)   shows the ratio $-2K/W$ versus the concentration for the haloes in the Magneticum Box0/mr run
while Figure \ref{fig:virial} (right panel)  shows $\eta$ versus the concentration.
The median $\eta$ is close to $0.9$ and it is generally lower than  the median of $-2K/W.$ 
Theoretical works as \cite{2016MNRAS.457.4340K} found a lower virial ratio when considering the term $E_s.$
From the figures we can see that there is a correlation between concentration and  $-2K/W,$ while the correlation is much weaker if we add $E_s$  to the kinetic term.

\begin{table}
\caption{Fit parameters of mass-concentration relation for relaxed, un-relaxed and all clusters performed   $z=0$  and with a pivot mass of $10^{14}\Msun$   }
 \begin{tabular}{l r r}
  \hline
 \multicolumn{3}{|c|}{Fit function $c_{200}=A\cdot\left(\frac{M_{200}}{10^{14}\Msun}\right)^B$}\\
 \hline 
  Halo samples & A & B\\
relaxed & $ 5.2  \pm 0.1$ & $-0.1 \pm 0.1$ \\
un-relaxed & $ 4.3  \pm 0.1$ & $-0.1 \pm 0.1$ \\
complete sample & $4.8  \pm 0.1$ & $ -0.1 \pm 0.1$ \\
  \hline 
  \end{tabular}
\label{table:relfit}
\end{table}

\nuovo{We identify un-relaxed clusters by selecting haloes with  $-2K/W$ lower than $0.5$ or greater than $1.5.$
Those objects have either a large imbalance between the total gravitational energy and the kinetic energy or a large energy from the surface pressure (and thus an inflow/outflow of material).
Table \ref{table:relfit} shows the fit performed at $z=0$ with the binning technique as for the previous fits of Eq. \ref{eq:fit_c}.
We used a pivot mass of $10^{14}\Msun$  in order to easily compare it with observations.}
 \nuovo{The values are in agreement with recent observations on SZE selected galaxy clusters \cite[see e.g. Table 7 in][]{2019MNRAS.482.1043C}.}
 
Figure \ref{fig:virialfossil} shows the fossil parameter as a function of $E_s/W$   colour coded by the concentration.
Fossil objects have lower $E_s$ (accreting less material from outside) than other clusters, thus their more external region has no activity (no in-fall or outfall of material).
This is also in agreement with  Figure  \ref{fig:c_grow} where the evolution of fossil concentration is dominated by their   internal motions (central galaxy accretes satellites).

\section{Conclusions}
\label{sec:concluz}

We used three cosmological hydrodynamic simulations from the Magneticum suite to cover a mass range  from $3\cdot10^{11}$ to $6\cdot10^{15}\Msun$ of well resolved clusters from redshift zero to redshift $2$ and we computed the concentration for all well resolved  haloes and  fit it as a power law of mass and redshift.

This is the first study of the mass-concentration relation in hydrodynamic simulations covering several orders of magnitude in mass. 
For high massive clusters, we found a value of the concentration and its dependency on mass and redshift is in agreement within the large scatter already present in both observations and simulations.

An exception is made for the low mass regime, wherein the Magneticum simulation  concentration \corr{(of the dark matter density)} is systematically lower than concentration found in studies based on dark matter only simulations.
\corr{Such different behaviour   is in agreement with other theoretical studies where the activation of  AGN feedback in low mass  haloes is capable of lowering the concentration up to a factor of $\approx15\%$   \citep[see Figure 8 in][]{2010MNRAS.405.2161D} by removing baryons from the inner region of the halo.
These effects have also been reproduced by the NIHAO  hydrodynamic cosmological simulations with high spatial resolution that reaches down to   $3\cdot10^3\Msun$ per gas particle.
\corr{\cite{2016MNRAS.462..663B}  find a flattening of the core region when comparing the dark matter density profile of a hydrodynamic run against its DMO counter part  and an overall decrease in $M_{200}$ \citep{2016MNRAS.461.2658D}.
In fact, the presence of baryons proved to be able to make dark matter haloes to be less cuspy \citep{2018arXiv181110625D}.}
These effects  contribute in lowering the concentration of   dark matter density profiles of low-mass haloes  (with respect to DMO runs). }
Thanks to the high mass regime of the Magneticum simulations we are able to  capture this effect and its  disappearance as the halo mass increases.

In the second part of this work we discussed the origin of the large scatter of concentration in the mass-concentration plane by studying its dependency on the fossilness.
Fossil groups are supposed to have had a long period of inactivity and are known to have a higher concentration \cite[see e.g.][]{2007MNRAS.381.1450N,2014MNRAS.441.3359D,2016A&A...590L...1P}.
Since we are working with hydrodynamic simulations, we compare halo fossilness (stellar mass ratio between central and most massive satellite of the system, as in see Eq. \ref{eq:fossilness})  with observations.
We find that the large statistics of Magneticum simulations is able to reproduce these rare objects.
Thanks to the large number of objects we are able to fit the concentration as a function of mass, redshift and fossil parameter (see Table \ref{table:fitfoss}), where we find a positive correlation between concentration and fossil parameter.

\corrd{We also investigate the underlying mechanism that brings fossil groups to the highest part of the mass-concentration plane.}
\corrd{For this reason we followed the time evolution of some haloes.}
Here we showed that in  unperturbed haloes, both fossilness and concentration steadily and slowly grow with time (see Figure \ref{fig:c_grow}).
This is in contrast with more naive models where an unperturbed halo keeps its concentration making it a mere function of its collapse time \citep[as in][]{2001MNRAS.321..559B}.
Interestingly, we found that this change of concentration is due to a decline of the scale radius.
We also showed how the scale radius   and  fossilness   increase or decrease together    when a major merger occurs (see Figure \ref{fig:c_event}).
From these analyses, we found that those two effects drive the correlation between concentration and fossil parameter.
Our findings are not in contrast with the fact that relaxed and fossil objects start with a high concentration because of their early formation times, but we show how an additional steady increase of the concentration pushes these objects in the very high region of the mass-concentration plane.

We then examined  the concentration as a function of the virial ratio $\eta=-(2K-E_s)/W$  and as a function of the energy from the surface pressure.
We found a weak dependency of the concentration on $-(2K-E_s)/W$ and very weak on the terms $-2K/W$ and $E_s.$
While a  large value of $E_s$ means that the cluster has a considerable amount of in-falling material and this translates into a low concentration and low fossil parameter; while a low value of $E_s$ (no in-falling material) can be related to both high and low concentrated clusters.
{The  difference between  $-2K/W$ and  $-\left(2K-E_s\right)/W$ is higher for haloes with lower  concentration.
This implies that low concentration objects are accreting material from the outside and it is in agreement with the idea that low-concentration haloes are not relaxed.}
This is compatible with other theoretical works as \cite{2016MNRAS.457.4340K}.
\corrd{This last analyses also showed that (see Figure  \ref{fig:virialfossil}) how fossil objects have both high concentration and a low value of $E_s,$ indicating a low accretion rate.
Our findings point to the direction that fossil objects lived un-perturbed, accreted all massive satellites  and have no in-fall/outfall material.}

Work has still to be done to study the relation between fossil parameter and other  quantities that are well known to be tied with the dynamical state of a system, for instance,  the difference between centre of mass and density peak position), or the velocity dispersion deviation between the one inferred from the virial theorem.
Additional work is also needed in order to understand the connection between central galaxy accreting satellites and the redistribution of the angular momentum within the halo, which in turn may give hints on the weak dependency between concentration and spin parameter \citep[as found by][]{2008MNRAS.391.1940M}. 

\section*{Acknowledgements}

The \textit{Magneticum Pathfinder} simulations were partially performed at the Leibniz-Rechenzentrum with CPU time assigned to the Project `pr86re'. This work was supported by the DFG Cluster of Excellence `Origin and Structure of the Universe'.
We are especially grateful for the support by M. Petkova through the Computational Center for Particle and Astrophysics (C$^2$PAP). Information on the \textit{Magneticum Pathfinder} project is available at \url{http://www.magneticum.org}.
Thanks to Rupam Bhattacharya for proof reading this manuscript, \corrd{Aura Obreja for some useful  references and the anonymous referee for requesting new details that improved the readability of this manuscript.}

\bibliographystyle{mnras}
\bibliography{auto}

\begin{thebibliography}{}
\makeatletter
\relax
\def\mn@urlcharsother{\let\do\@makeother \do\$\do\&\do\#\do\^\do\_\do\%\do\~}
\def\mn@doi{\begingroup\mn@urlcharsother \@ifnextchar [ {\mn@doi@}
  {\mn@doi@[]}}
\def\mn@doi@[#1]#2{\def\@tempa{#1}\ifx\@tempa\@empty \href
  {http://dx.doi.org/#2} {doi:#2}\else \href {http://dx.doi.org/#2} {#1}\fi
  \endgroup}
\def\mn@eprint#1#2{\mn@eprint@#1:#2::\@nil}
\def\mn@eprint@arXiv#1{\href {http://arxiv.org/abs/#1} {{\tt arXiv:#1}}}
\def\mn@eprint@dblp#1{\href {http://dblp.uni-trier.de/rec/bibtex/#1.xml}
  {dblp:#1}}
\def\mn@eprint@#1:#2:#3:#4\@nil{\def\@tempa {#1}\def\@tempb {#2}\def\@tempc
  {#3}\ifx \@tempc \@empty \let \@tempc \@tempb \let \@tempb \@tempa \fi \ifx
  \@tempb \@empty \def\@tempb {arXiv}\fi \@ifundefined
  {mn@eprint@\@tempb}{\@tempb:\@tempc}{\expandafter \expandafter \csname
  mn@eprint@\@tempb\endcsname \expandafter{\@tempc}}}

\bibitem[\protect\citeauthoryear{{Bartalucci}, {Arnaud}, {Pratt}  \& {Le
  Brun}}{{Bartalucci} et~al.}{2018}]{2018A&A...617A..64B}
{Bartalucci} I.,  {Arnaud} M.,  {Pratt} G.~W.,   {Le Brun} A.~M.~C.,  2018,
  \mn@doi [\aap] {10.1051/0004-6361/201732458}, \href
  {http://adsabs.harvard.edu/abs/2018A%26A...617A..64B} {617, A64}

\bibitem[\protect\citeauthoryear{{Beck} et~al.,}{{Beck}
  et~al.}{2016}]{2016MNRAS.455.2110B}
{Beck} A.~M.,  et~al., 2016, \mn@doi [\mnras] {10.1093/mnras/stv2443}, \href
  {http://adsabs.harvard.edu/abs/2016MNRAS.455.2110B} {455, 2110}

\bibitem[\protect\citeauthoryear{{Bellstedt} et~al.,}{{Bellstedt}
  et~al.}{2018}]{2018MNRAS.476.4543B}
{Bellstedt} S.,  et~al., 2018, \mn@doi [\mnras] {10.1093/mnras/sty456}, \href
  {http://adsabs.harvard.edu/abs/2018MNRAS.476.4543B} {476, 4543}

\bibitem[\protect\citeauthoryear{{Bhattacharya}, {Habib}, {Heitmann}  \&
  {Vikhlinin}}{{Bhattacharya} et~al.}{2013}]{2013ApJ...766...32B}
{Bhattacharya} S.,  {Habib} S.,  {Heitmann} K.,   {Vikhlinin} A.,  2013,
  \mn@doi [\apj] {10.1088/0004-637X/766/1/32}, \href
  {http://adsabs.harvard.edu/abs/2013ApJ...766...32B} {766, 32}

\bibitem[\protect\citeauthoryear{{Biffi}, {Dolag}  \& {B{\"o}hringer}}{{Biffi}
  et~al.}{2013}]{2013MNRAS.428.1395B}
{Biffi} V.,  {Dolag} K.,   {B{\"o}hringer} H.,  2013, \mn@doi [\mnras]
  {10.1093/mnras/sts120}, \href
  {http://adsabs.harvard.edu/abs/2013MNRAS.428.1395B} {428, 1395}

\bibitem[\protect\citeauthoryear{{Binney} \& {Tremaine}}{{Binney} \&
  {Tremaine}}{2008}]{2008gady.book.....B}
{Binney} J.,  {Tremaine} S.,  2008, {Galactic Dynamics: Second Edition}.
Princeton University Press

\bibitem[\protect\citeauthoryear{{Biviano} et~al.,}{{Biviano}
  et~al.}{2017}]{2017A&A...607A..81B}
{Biviano} A.,  et~al., 2017, \mn@doi [\aap] {10.1051/0004-6361/201731289},
  \href {http://adsabs.harvard.edu/abs/2017A%26A...607A..81B} {607, A81}

\bibitem[\protect\citeauthoryear{{Bocquet}, {Saro}, {Dolag}  \&
  {Mohr}}{{Bocquet} et~al.}{2016}]{2016MNRAS.456.2361B}
{Bocquet} S.,  {Saro} A.,  {Dolag} K.,   {Mohr} J.~J.,  2016, \mn@doi [\mnras]
  {10.1093/mnras/stv2657}, \href
  {http://adsabs.harvard.edu/abs/2016MNRAS.456.2361B} {456, 2361}

\bibitem[\protect\citeauthoryear{{Borgani} \& {Kravtsov}}{{Borgani} \&
  {Kravtsov}}{2011}]{2011ASL.....4..204B}
{Borgani} S.,  {Kravtsov} A.,  2011, \mn@doi [Advanced Science Letters]
  {10.1166/asl.2011.1209}, \href
  {http://adsabs.harvard.edu/abs/2011ASL.....4..204B} {4, 204}

\bibitem[\protect\citeauthoryear{{Bullock}, {Kolatt}, {Sigad}, {Somerville},
  {Kravtsov}, {Klypin}, {Primack}  \& {Dekel}}{{Bullock}
  et~al.}{2001}]{2001MNRAS.321..559B}
{Bullock} J.~S.,  {Kolatt} T.~S.,  {Sigad} Y.,  {Somerville} R.~S.,  {Kravtsov}
  A.~V.,  {Klypin} A.~A.,  {Primack} J.~R.,   {Dekel} A.,  2001, \mn@doi
  [\mnras] {10.1046/j.1365-8711.2001.04068.x}, \href
  {http://adsabs.harvard.edu/abs/2001MNRAS.321..559B} {321, 559}

\bibitem[\protect\citeauthoryear{{Buote}}{{Buote}}{2017}]{2017ApJ...834..164B}
{Buote} D.~A.,  2017, \mn@doi [\apj] {10.3847/1538-4357/834/2/164}, \href
  {http://adsabs.harvard.edu/abs/2017ApJ...834..164B} {834, 164}

\bibitem[\protect\citeauthoryear{{Butsky} et~al.,}{{Butsky}
  et~al.}{2016}]{2016MNRAS.462..663B}
{Butsky} I.,  et~al., 2016, \mn@doi [\mnras] {10.1093/mnras/stw1688}, \href
  {http://adsabs.harvard.edu/abs/2016MNRAS.462..663B} {462, 663}

\bibitem[\protect\citeauthoryear{{Capasso} et~al.,}{{Capasso}
  et~al.}{2019}]{2019MNRAS.482.1043C}
{Capasso} R.,  et~al., 2019, \mn@doi [\mnras] {10.1093/mnras/sty2645}, \href
  {http://adsabs.harvard.edu/abs/2019MNRAS.482.1043C} {482, 1043}

\bibitem[\protect\citeauthoryear{{Chan}, {Kere{\v s}}, {Wetzel}, {Hopkins},
  {Faucher-Gigu{\`e}re}, {El-Badry}, {Garrison-Kimmel}  \&
  {Boylan-Kolchin}}{{Chan} et~al.}{2018}]{2018MNRAS.478..906C}
{Chan} T.~K.,  {Kere{\v s}} D.,  {Wetzel} A.,  {Hopkins} P.~F.,
  {Faucher-Gigu{\`e}re} C.-A.,  {El-Badry} K.,  {Garrison-Kimmel} S.,
  {Boylan-Kolchin} M.,  2018, \mn@doi [\mnras] {10.1093/mnras/sty1153}, \href
  {http://adsabs.harvard.edu/abs/2018MNRAS.478..906C} {478, 906}

\bibitem[\protect\citeauthoryear{{Chandrasekhar}}{{Chandrasekhar}}{1961}]{1961hhs..book.....C}
{Chandrasekhar} S.,  1961, {Hydrodynamic and hydromagnetic stability}

\bibitem[\protect\citeauthoryear{{Coe} et~al.,}{{Coe}
  et~al.}{2012}]{2012ApJ...757...22C}
{Coe} D.,  et~al., 2012, \mn@doi [\apj] {10.1088/0004-637X/757/1/22}, \href
  {http://adsabs.harvard.edu/abs/2012ApJ...757...22C} {757, 22}

\bibitem[\protect\citeauthoryear{{Correa}, {Wyithe}, {Schaye}  \&
  {Duffy}}{{Correa} et~al.}{2015}]{2015MNRAS.452.1217C}
{Correa} C.~A.,  {Wyithe} J.~S.~B.,  {Schaye} J.,   {Duffy} A.~R.,  2015,
  \mn@doi [\mnras] {10.1093/mnras/stv1363}, \href
  {http://adsabs.harvard.edu/abs/2015MNRAS.452.1217C} {452, 1217}

\bibitem[\protect\citeauthoryear{{Corsini} et~al.,}{{Corsini}
  et~al.}{2018}]{2018A&A...618A.172C}
{Corsini} E.~M.,  et~al., 2018, \mn@doi [\aap] {10.1051/0004-6361/201832625},
  \href {http://adsabs.harvard.edu/abs/2018A%26A...618A.172C} {618, A172}

\bibitem[\protect\citeauthoryear{{Covone}, {Sereno}, {Kilbinger}  \&
  {Cardone}}{{Covone} et~al.}{2014}]{2014ApJ...784L..25C}
{Covone} G.,  {Sereno} M.,  {Kilbinger} M.,   {Cardone} V.~F.,  2014, \mn@doi
  [\apjl] {10.1088/2041-8205/784/2/L25}, \href
  {http://adsabs.harvard.edu/abs/2014ApJ...784L..25C} {784, L25}

\bibitem[\protect\citeauthoryear{{Cui}, {Power}, {Borgani}, {Knebe}, {Lewis},
  {Murante}  \& {Poole}}{{Cui} et~al.}{2017}]{2017MNRAS.464.2502C}
{Cui} W.,  {Power} C.,  {Borgani} S.,  {Knebe} A.,  {Lewis} G.~F.,  {Murante}
  G.,   {Poole} G.~B.,  2017, \mn@doi [\mnras] {10.1093/mnras/stw2567}, \href
  {http://adsabs.harvard.edu/abs/2017MNRAS.464.2502C} {464, 2502}

\bibitem[\protect\citeauthoryear{{De Boni}}{{De
  Boni}}{2013}]{2013arXiv1302.2364D}
{De Boni} C.,  2013, arXiv e-prints, \href
  {http://adsabs.harvard.edu/abs/2013arXiv1302.2364D} {}

\bibitem[\protect\citeauthoryear{{De Boni}, {Ettori}, {Dolag}  \&
  {Moscardini}}{{De Boni} et~al.}{2013}]{2013MNRAS.428.2921D}
{De Boni} C.,  {Ettori} S.,  {Dolag} K.,   {Moscardini} L.,  2013, \mn@doi
  [\mnras] {10.1093/mnras/sts235}, \href
  {http://adsabs.harvard.edu/abs/2013MNRAS.428.2921D} {428, 2921}

\bibitem[\protect\citeauthoryear{{Dolag}, {Bartelmann}, {Perrotta},
  {Baccigalupi}, {Moscardini}, {Meneghetti}  \& {Tormen}}{{Dolag}
  et~al.}{2004}]{2004A&A...416..853D}
{Dolag} K.,  {Bartelmann} M.,  {Perrotta} F.,  {Baccigalupi} C.,  {Moscardini}
  L.,  {Meneghetti} M.,   {Tormen} G.,  2004, \mn@doi [\aap]
  {10.1051/0004-6361:20031757}, \href
  {http://adsabs.harvard.edu/abs/2004A%26A...416..853D} {416, 853}

\bibitem[\protect\citeauthoryear{{Dolag}, {Borgani}, {Murante}  \&
  {Springel}}{{Dolag} et~al.}{2009}]{2009MNRAS.399..497D}
{Dolag} K.,  {Borgani} S.,  {Murante} G.,   {Springel} V.,  2009, \mn@doi
  [\mnras] {10.1111/j.1365-2966.2009.15034.x}, \href
  {http://adsabs.harvard.edu/abs/2009MNRAS.399..497D} {399, 497}

\bibitem[\protect\citeauthoryear{{Dolag}, {Gaensler}, {Beck}  \&
  {Beck}}{{Dolag} et~al.}{2015}]{2015MNRAS.451.4277D}
{Dolag} K.,  {Gaensler} B.~M.,  {Beck} A.~M.,   {Beck} M.~C.,  2015, \mn@doi
  [\mnras] {10.1093/mnras/stv1190}, \href
  {http://adsabs.harvard.edu/abs/2015MNRAS.451.4277D} {451, 4277}

\bibitem[\protect\citeauthoryear{{Dolag}, {Komatsu}  \& {Sunyaev}}{{Dolag}
  et~al.}{2016}]{2016MNRAS.463.1797D}
{Dolag} K.,  {Komatsu} E.,   {Sunyaev} R.,  2016, \mn@doi [\mnras]
  {10.1093/mnras/stw2035}, \href
  {http://adsabs.harvard.edu/abs/2016MNRAS.463.1797D} {463, 1797}

\bibitem[\protect\citeauthoryear{{Du}, {Fan}, {Shan}, {Zhao}, {Covone}, {Fu}
  \& {Kneib}}{{Du} et~al.}{2015}]{2015ApJ...814..120D}
{Du} W.,  {Fan} Z.,  {Shan} H.,  {Zhao} G.-B.,  {Covone} G.,  {Fu} L.,
  {Kneib} J.-P.,  2015, \mn@doi [\apj] {10.1088/0004-637X/814/2/120}, \href
  {http://adsabs.harvard.edu/abs/2015ApJ...814..120D} {814, 120}

\bibitem[\protect\citeauthoryear{{Duffy}, {Schaye}, {Kay}  \& {Dalla
  Vecchia}}{{Duffy} et~al.}{2008}]{2008MNRAS.390L..64D}
{Duffy} A.~R.,  {Schaye} J.,  {Kay} S.~T.,   {Dalla Vecchia} C.,  2008, \mn@doi
  [\mnras] {10.1111/j.1745-3933.2008.00537.x}, \href
  {http://adsabs.harvard.edu/abs/2008MNRAS.390L..64D} {390, L64}

\bibitem[\protect\citeauthoryear{{Duffy}, {Schaye}, {Kay}, {Dalla Vecchia},
  {Battye}  \& {Booth}}{{Duffy} et~al.}{2010}]{2010MNRAS.405.2161D}
{Duffy} A.~R.,  {Schaye} J.,  {Kay} S.~T.,  {Dalla Vecchia} C.,  {Battye}
  R.~A.,   {Booth} C.~M.,  2010, \mn@doi [\mnras]
  {10.1111/j.1365-2966.2010.16613.x}, \href
  {http://adsabs.harvard.edu/abs/2010MNRAS.405.2161D} {405, 2161}

\bibitem[\protect\citeauthoryear{{Dutton} \& {Macci{\`o}}}{{Dutton} \&
  {Macci{\`o}}}{2014}]{2014MNRAS.441.3359D}
{Dutton} A.~A.,  {Macci{\`o}} A.~V.,  2014, \mn@doi [\mnras]
  {10.1093/mnras/stu742}, \href
  {http://adsabs.harvard.edu/abs/2014MNRAS.441.3359D} {441, 3359}

\bibitem[\protect\citeauthoryear{{Dutton} et~al.,}{{Dutton}
  et~al.}{2016}]{2016MNRAS.461.2658D}
{Dutton} A.~A.,  et~al., 2016, \mn@doi [\mnras] {10.1093/mnras/stw1537}, \href
  {http://adsabs.harvard.edu/abs/2016MNRAS.461.2658D} {461, 2658}

\bibitem[\protect\citeauthoryear{{Dutton}, {Macci{\`o}}, {Buck}, {Dixon},
  {Blank}  \& {Obreja}}{{Dutton} et~al.}{2018}]{2018arXiv181110625D}
{Dutton} A.~A.,  {Macci{\`o}} A.~V.,  {Buck} T.,  {Dixon} K.~L.,  {Blank} M.,
  {Obreja} A.,  2018, arXiv e-prints, \href
  {http://adsabs.harvard.edu/abs/2018arXiv181110625D} {}

\bibitem[\protect\citeauthoryear{{El-Badry}, {Wetzel}, {Geha}, {Hopkins},
  {Kere{\v s}}, {Chan}  \& {Faucher-Gigu{\`e}re}}{{El-Badry}
  et~al.}{2016}]{2016ApJ...820..131E}
{El-Badry} K.,  {Wetzel} A.,  {Geha} M.,  {Hopkins} P.~F.,  {Kere{\v s}} D.,
  {Chan} T.~K.,   {Faucher-Gigu{\`e}re} C.-A.,  2016, \mn@doi [\apj]
  {10.3847/0004-637X/820/2/131}, \href
  {http://adsabs.harvard.edu/abs/2016ApJ...820..131E} {820, 131}

\bibitem[\protect\citeauthoryear{{Fabjan}, {Borgani}, {Tornatore}, {Saro},
  {Murante}  \& {Dolag}}{{Fabjan} et~al.}{2010}]{2010MNRAS.401.1670F}
{Fabjan} D.,  {Borgani} S.,  {Tornatore} L.,  {Saro} A.,  {Murante} G.,
  {Dolag} K.,  2010, \mn@doi [\mnras] {10.1111/j.1365-2966.2009.15794.x}, \href
  {http://adsabs.harvard.edu/abs/2010MNRAS.401.1670F} {401, 1670}

\bibitem[\protect\citeauthoryear{{Ferland}, {Korista}, {Verner}, {Ferguson},
  {Kingdon}  \& {Verner}}{{Ferland} et~al.}{1998}]{1998PASP..110..761F}
{Ferland} G.~J.,  {Korista} K.~T.,  {Verner} D.~A.,  {Ferguson} J.~W.,
  {Kingdon} J.~B.,   {Verner} E.~M.,  1998, \mn@doi [\pasp] {10.1086/316190},
  \href {http://adsabs.harvard.edu/abs/1998PASP..110..761F} {110, 761}

\bibitem[\protect\citeauthoryear{{Fujita}, {Umetsu}, {Rasia}, {Meneghetti},
  {Donahue}, {Medezinski}, {Okabe}  \& {Postman}}{{Fujita}
  et~al.}{2018a}]{2018ApJ...857..118F}
{Fujita} Y.,  {Umetsu} K.,  {Rasia} E.,  {Meneghetti} M.,  {Donahue} M.,
  {Medezinski} E.,  {Okabe} N.,   {Postman} M.,  2018a, \mn@doi [\apj]
  {10.3847/1538-4357/aab8fd}, \href
  {http://adsabs.harvard.edu/abs/2018ApJ...857..118F} {857, 118}

\bibitem[\protect\citeauthoryear{{Fujita}, {Umetsu}, {Ettori}, {Rasia}, {Okabe}
   \& {Meneghetti}}{{Fujita} et~al.}{2018b}]{2018ApJ...863...37F}
{Fujita} Y.,  {Umetsu} K.,  {Ettori} S.,  {Rasia} E.,  {Okabe} N.,
  {Meneghetti} M.,  2018b, \mn@doi [\apj] {10.3847/1538-4357/aacf05}, \href
  {http://adsabs.harvard.edu/abs/2018ApJ...863...37F} {863, 37}

\bibitem[\protect\citeauthoryear{{Giocoli}, {Meneghetti}, {Ettori}  \&
  {Moscardini}}{{Giocoli} et~al.}{2012}]{2012MNRAS.426.1558G}
{Giocoli} C.,  {Meneghetti} M.,  {Ettori} S.,   {Moscardini} L.,  2012, \mn@doi
  [\mnras] {10.1111/j.1365-2966.2012.21743.x}, \href
  {http://adsabs.harvard.edu/abs/2012MNRAS.426.1558G} {426, 1558}

\bibitem[\protect\citeauthoryear{{Groener}, {Goldberg}  \& {Sereno}}{{Groener}
  et~al.}{2016}]{2016MNRAS.455..892G}
{Groener} A.~M.,  {Goldberg} D.~M.,   {Sereno} M.,  2016, \mn@doi [\mnras]
  {10.1093/mnras/stv2341}, \href
  {http://adsabs.harvard.edu/abs/2016MNRAS.455..892G} {455, 892}

\bibitem[\protect\citeauthoryear{{Gunn} \& {Gott}}{{Gunn} \&
  {Gott}}{1972}]{1972ApJ...176....1G}
{Gunn} J.~E.,  {Gott} III J.~R.,  1972, \mn@doi [\apj] {10.1086/151605}, \href
  {http://adsabs.harvard.edu/abs/1972ApJ...176....1G} {176, 1}

\bibitem[\protect\citeauthoryear{{Hirschmann}, {Dolag}, {Saro}, {Bachmann},
  {Borgani}  \& {Burkert}}{{Hirschmann} et~al.}{2014}]{2014MNRAS.442.2304H}
{Hirschmann} M.,  {Dolag} K.,  {Saro} A.,  {Bachmann} L.,  {Borgani} S.,
  {Burkert} A.,  2014, \mn@doi [\mnras] {10.1093/mnras/stu1023}, \href
  {http://adsabs.harvard.edu/abs/2014MNRAS.442.2304H} {442, 2304}

\bibitem[\protect\citeauthoryear{{Humphrey}, {Buote}, {Canizares}, {Fabian}  \&
  {Miller}}{{Humphrey} et~al.}{2011}]{2011ApJ...729...53H}
{Humphrey} P.~J.,  {Buote} D.~A.,  {Canizares} C.~R.,  {Fabian} A.~C.,
  {Miller} J.~M.,  2011, \mn@doi [\apj] {10.1088/0004-637X/729/1/53}, \href
  {http://adsabs.harvard.edu/abs/2011ApJ...729...53H} {729, 53}

\bibitem[\protect\citeauthoryear{{Humphrey}, {Buote}, {O'Sullivan}  \&
  {Ponman}}{{Humphrey} et~al.}{2012}]{2012ApJ...755..166H}
{Humphrey} P.~J.,  {Buote} D.~A.,  {O'Sullivan} E.,   {Ponman} T.~J.,  2012,
  \mn@doi [\apj] {10.1088/0004-637X/755/2/166}, \href
  {http://adsabs.harvard.edu/abs/2012ApJ...755..166H} {755, 166}

\bibitem[\protect\citeauthoryear{{Khosroshahi}, {Maughan}, {Ponman}  \&
  {Jones}}{{Khosroshahi} et~al.}{2006}]{2006MNRAS.369.1211K}
{Khosroshahi} H.~G.,  {Maughan} B.~J.,  {Ponman} T.~J.,   {Jones} L.~R.,  2006,
  \mn@doi [\mnras] {10.1111/j.1365-2966.2006.10357.x}, \href
  {http://adsabs.harvard.edu/abs/2006MNRAS.369.1211K} {369, 1211}

\bibitem[\protect\citeauthoryear{{Klypin}, {Trujillo-Gomez}  \&
  {Primack}}{{Klypin} et~al.}{2011}]{2011ApJ...740..102K}
{Klypin} A.~A.,  {Trujillo-Gomez} S.,   {Primack} J.,  2011, \mn@doi [\apj]
  {10.1088/0004-637X/740/2/102}, \href
  {http://adsabs.harvard.edu/abs/2011ApJ...740..102K} {740, 102}

\bibitem[\protect\citeauthoryear{{Klypin}, {Yepes}, {Gottl{\"o}ber}, {Prada}
  \& {He{\ss}}}{{Klypin} et~al.}{2016}]{2016MNRAS.457.4340K}
{Klypin} A.,  {Yepes} G.,  {Gottl{\"o}ber} S.,  {Prada} F.,   {He{\ss}} S.,
  2016, \mn@doi [\mnras] {10.1093/mnras/stw248}, \href
  {http://adsabs.harvard.edu/abs/2016MNRAS.457.4340K} {457, 4340}

\bibitem[\protect\citeauthoryear{{Komatsu} et~al.,}{{Komatsu}
  et~al.}{2011}]{2011ApJS..192...18K}
{Komatsu} E.,  et~al., 2011, \mn@doi [\apjs] {10.1088/0067-0049/192/2/18},
  \href {http://adsabs.harvard.edu/abs/2011ApJS..192...18K} {192, 18}

\bibitem[\protect\citeauthoryear{{Kundert} et~al.,}{{Kundert}
  et~al.}{2015}]{2015MNRAS.454..161K}
{Kundert} A.,  et~al., 2015, \mn@doi [\mnras] {10.1093/mnras/stv1879}, \href
  {http://adsabs.harvard.edu/abs/2015MNRAS.454..161K} {454, 161}

\bibitem[\protect\citeauthoryear{{Lin}, {Jing}, {Mao}, {Gao}  \&
  {McCarthy}}{{Lin} et~al.}{2006}]{2006ApJ...651..636L}
{Lin} W.~P.,  {Jing} Y.~P.,  {Mao} S.,  {Gao} L.,   {McCarthy} I.~G.,  2006,
  \mn@doi [\apj] {10.1086/508052}, \href
  {http://adsabs.harvard.edu/abs/2006ApJ...651..636L} {651, 636}

\bibitem[\protect\citeauthoryear{{Ludlow}, {Navarro}, {Li}, {Angulo},
  {Boylan-Kolchin}  \& {Bett}}{{Ludlow} et~al.}{2012}]{2012MNRAS.427.1322L}
{Ludlow} A.~D.,  {Navarro} J.~F.,  {Li} M.,  {Angulo} R.~E.,  {Boylan-Kolchin}
  M.,   {Bett} P.~E.,  2012, \mn@doi [\mnras]
  {10.1111/j.1365-2966.2012.21892.x}, \href
  {http://adsabs.harvard.edu/abs/2012MNRAS.427.1322L} {427, 1322}

\bibitem[\protect\citeauthoryear{{Ludlow}, {Navarro}, {Angulo},
  {Boylan-Kolchin}, {Springel}, {Frenk}  \& {White}}{{Ludlow}
  et~al.}{2014}]{2014MNRAS.441..378L}
{Ludlow} A.~D.,  {Navarro} J.~F.,  {Angulo} R.~E.,  {Boylan-Kolchin} M.,
  {Springel} V.,  {Frenk} C.,   {White} S.~D.~M.,  2014, \mn@doi [\mnras]
  {10.1093/mnras/stu483}, \href
  {http://adsabs.harvard.edu/abs/2014MNRAS.441..378L} {441, 378}

\bibitem[\protect\citeauthoryear{{Macci{\`o}}, {Dutton}, {van den Bosch},
  {Moore}, {Potter}  \& {Stadel}}{{Macci{\`o}}
  et~al.}{2007}]{2007MNRAS.378...55M}
{Macci{\`o}} A.~V.,  {Dutton} A.~A.,  {van den Bosch} F.~C.,  {Moore} B.,
  {Potter} D.,   {Stadel} J.,  2007, \mn@doi [\mnras]
  {10.1111/j.1365-2966.2007.11720.x}, \href
  {http://adsabs.harvard.edu/abs/2007MNRAS.378...55M} {378, 55}

\bibitem[\protect\citeauthoryear{{Macci{\`o}}, {Dutton}  \& {van den
  Bosch}}{{Macci{\`o}} et~al.}{2008}]{2008MNRAS.391.1940M}
{Macci{\`o}} A.~V.,  {Dutton} A.~A.,   {van den Bosch} F.~C.,  2008, \mn@doi
  [\mnras] {10.1111/j.1365-2966.2008.14029.x}, \href
  {http://adsabs.harvard.edu/abs/2008MNRAS.391.1940M} {391, 1940}

\bibitem[\protect\citeauthoryear{{Mandelbaum}, {Seljak}  \&
  {Hirata}}{{Mandelbaum} et~al.}{2008}]{2008JCAP...08..006M}
{Mandelbaum} R.,  {Seljak} U.,   {Hirata} C.~M.,  2008, \mn@doi [\jcap]
  {10.1088/1475-7516/2008/08/006}, \href
  {http://adsabs.harvard.edu/abs/2008JCAP...08..006M} {8, 006}

\bibitem[\protect\citeauthoryear{{Mantz}, {Allen}  \& {Morris}}{{Mantz}
  et~al.}{2016}]{2016MNRAS.462..681M}
{Mantz} A.~B.,  {Allen} S.~W.,   {Morris} R.~G.,  2016, \mn@doi [\mnras]
  {10.1093/mnras/stw1707}, \href
  {http://adsabs.harvard.edu/abs/2016MNRAS.462..681M} {462, 681}

\bibitem[\protect\citeauthoryear{{Martinsson}, {Verheijen}, {Westfall},
  {Bershady}, {Andersen}  \& {Swaters}}{{Martinsson}
  et~al.}{2013}]{2013A&A...557A.131M}
{Martinsson} T.~P.~K.,  {Verheijen} M.~A.~W.,  {Westfall} K.~B.,  {Bershady}
  M.~A.,  {Andersen} D.~R.,   {Swaters} R.~A.,  2013, \mn@doi [\aap]
  {10.1051/0004-6361/201321390}, \href
  {http://adsabs.harvard.edu/abs/2013A%26A...557A.131M} {557, A131}

\bibitem[\protect\citeauthoryear{{Meneghetti} \& {Rasia}}{{Meneghetti} \&
  {Rasia}}{2013}]{2013arXiv1303.6158M}
{Meneghetti} M.,  {Rasia} E.,  2013, arXiv e-prints, \href
  {http://adsabs.harvard.edu/abs/2013arXiv1303.6158M} {}

\bibitem[\protect\citeauthoryear{{Meneghetti}, {Argazzi}, {Pace}, {Moscardini},
  {Dolag}, {Bartelmann}, {Li}  \& {Oguri}}{{Meneghetti}
  et~al.}{2007}]{2007A&A...461...25M}
{Meneghetti} M.,  {Argazzi} R.,  {Pace} F.,  {Moscardini} L.,  {Dolag} K.,
  {Bartelmann} M.,  {Li} G.,   {Oguri} M.,  2007, \mn@doi [\aap]
  {10.1051/0004-6361:20065722}, \href
  {http://adsabs.harvard.edu/abs/2007A%26A...461...25M} {461, 25}

\bibitem[\protect\citeauthoryear{{Meneghetti} et~al.,}{{Meneghetti}
  et~al.}{2014}]{2014ApJ...797...34M}
{Meneghetti} M.,  et~al., 2014, \mn@doi [\apj] {10.1088/0004-637X/797/1/34},
  \href {http://adsabs.harvard.edu/abs/2014ApJ...797...34M} {797, 34}

\bibitem[\protect\citeauthoryear{{Merten} et~al.,}{{Merten}
  et~al.}{2015}]{2015ApJ...806....4M}
{Merten} J.,  et~al., 2015, \mn@doi [\apj] {10.1088/0004-637X/806/1/4}, \href
  {http://adsabs.harvard.edu/abs/2015ApJ...806....4M} {806, 4}

\bibitem[\protect\citeauthoryear{{Moore}, {Governato}, {Quinn}, {Stadel}  \&
  {Lake}}{{Moore} et~al.}{1998}]{1998ApJ...499L...5M}
{Moore} B.,  {Governato} F.,  {Quinn} T.,  {Stadel} J.,   {Lake} G.,  1998,
  \mn@doi [\apjl] {10.1086/311333}, \href
  {http://adsabs.harvard.edu/abs/1998ApJ...499L...5M} {499, L5}

\bibitem[\protect\citeauthoryear{{Naderi}, {Malekjani}  \& {Pace}}{{Naderi}
  et~al.}{2015}]{2015MNRAS.447.1873N}
{Naderi} T.,  {Malekjani} M.,   {Pace} F.,  2015, \mn@doi [\mnras]
  {10.1093/mnras/stu2481}, \href
  {http://adsabs.harvard.edu/abs/2015MNRAS.447.1873N} {447, 1873}

\bibitem[\protect\citeauthoryear{{Navarro}, {Frenk}  \& {White}}{{Navarro}
  et~al.}{1996}]{1996ApJ...462..563N}
{Navarro} J.~F.,  {Frenk} C.~S.,   {White} S.~D.~M.,  1996, \mn@doi [\apj]
  {10.1086/177173}, \href {http://adsabs.harvard.edu/abs/1996ApJ...462..563N}
  {462, 563}

\bibitem[\protect\citeauthoryear{{Navarro}, {Frenk}  \& {White}}{{Navarro}
  et~al.}{1997}]{1997ApJ...490..493N}
{Navarro} J.~F.,  {Frenk} C.~S.,   {White} S.~D.~M.,  1997, \mn@doi [\apj]
  {10.1086/304888}, \href {http://adsabs.harvard.edu/abs/1997ApJ...490..493N}
  {490, 493}

\bibitem[\protect\citeauthoryear{{Neto} et~al.,}{{Neto}
  et~al.}{2007}]{2007MNRAS.381.1450N}
{Neto} A.~F.,  et~al., 2007, \mn@doi [\mnras]
  {10.1111/j.1365-2966.2007.12381.x}, \href
  {http://adsabs.harvard.edu/abs/2007MNRAS.381.1450N} {381, 1450}

\bibitem[\protect\citeauthoryear{{Prada}, {Klypin}, {Cuesta}, {Betancort-Rijo}
  \& {Primack}}{{Prada} et~al.}{2012}]{2012MNRAS.423.3018P}
{Prada} F.,  {Klypin} A.~A.,  {Cuesta} A.~J.,  {Betancort-Rijo} J.~E.,
  {Primack} J.,  2012, \mn@doi [\mnras] {10.1111/j.1365-2966.2012.21007.x},
  \href {http://adsabs.harvard.edu/abs/2012MNRAS.423.3018P} {423, 3018}

\bibitem[\protect\citeauthoryear{{Pratt} \& {Arnaud}}{{Pratt} \&
  {Arnaud}}{2005}]{2005A&A...429..791P}
{Pratt} G.~W.,  {Arnaud} M.,  2005, \mn@doi [\aap]
  {10.1051/0004-6361:20041479}, \href
  {http://adsabs.harvard.edu/abs/2005A%26A...429..791P} {429, 791}

\bibitem[\protect\citeauthoryear{{Pratt}, {Pointecouteau}, {Arnaud}  \& {van
  der Burg}}{{Pratt} et~al.}{2016}]{2016A&A...590L...1P}
{Pratt} G.~W.,  {Pointecouteau} E.,  {Arnaud} M.,   {van der Burg} R.~F.~J.,
  2016, \mn@doi [\aap] {10.1051/0004-6361/201628462}, \href
  {http://adsabs.harvard.edu/abs/2016A%26A...590L...1P} {590, L1}

\bibitem[\protect\citeauthoryear{{Remus} \& {Dolag}}{{Remus} \&
  {Dolag}}{2016}]{2016ilgp.confE..43R}
{Remus} R.-S.,  {Dolag} K.,  2016, in The Interplay between Local and Global
  Processes in Galaxies,. p.~43

\bibitem[\protect\citeauthoryear{{Remus}, {Dolag}, {Naab}, {Burkert},
  {Hirschmann}, {Hoffmann}  \& {Johansson}}{{Remus}
  et~al.}{2017}]{2017MNRAS.464.3742R}
{Remus} R.-S.,  {Dolag} K.,  {Naab} T.,  {Burkert} A.,  {Hirschmann} M.,
  {Hoffmann} T.~L.,   {Johansson} P.~H.,  2017, \mn@doi [\mnras]
  {10.1093/mnras/stw2594}, \href
  {http://adsabs.harvard.edu/abs/2017MNRAS.464.3742R} {464, 3742}

\bibitem[\protect\citeauthoryear{{Rey}, {Pontzen}  \& {Saintonge}}{{Rey}
  et~al.}{2018}]{2018arXiv181009473R}
{Rey} M.~P.,  {Pontzen} A.,   {Saintonge} A.,  2018, arXiv e-prints, \href
  {http://adsabs.harvard.edu/abs/2018arXiv181009473R} {}

\bibitem[\protect\citeauthoryear{{Saro} et~al.,}{{Saro}
  et~al.}{2014}]{2014MNRAS.440.2610S}
{Saro} A.,  et~al., 2014, \mn@doi [\mnras] {10.1093/mnras/stu575}, \href
  {http://adsabs.harvard.edu/abs/2014MNRAS.440.2610S} {440, 2610}

\bibitem[\protect\citeauthoryear{{Schulze}, {Remus}, {Dolag}, {Burkert},
  {Emsellem}  \& {van de Ven}}{{Schulze} et~al.}{2018}]{2018MNRAS.480.4636S}
{Schulze} F.,  {Remus} R.-S.,  {Dolag} K.,  {Burkert} A.,  {Emsellem} E.,
  {van de Ven} G.,  2018, \mn@doi [\mnras] {10.1093/mnras/sty2090}, \href
  {http://adsabs.harvard.edu/abs/2018MNRAS.480.4636S} {480, 4636}

\bibitem[\protect\citeauthoryear{{Shan} et~al.,}{{Shan}
  et~al.}{2017}]{2017ApJ...840..104S}
{Shan} H.,  et~al., 2017, \mn@doi [\apj] {10.3847/1538-4357/aa6c68}, \href
  {http://adsabs.harvard.edu/abs/2017ApJ...840..104S} {840, 104}

\bibitem[\protect\citeauthoryear{{Shirasaki}, {Lau}  \& {Nagai}}{{Shirasaki}
  et~al.}{2018}]{2018MNRAS.477.2804S}
{Shirasaki} M.,  {Lau} E.~T.,   {Nagai} D.,  2018, \mn@doi [\mnras]
  {10.1093/mnras/sty763}, \href
  {http://adsabs.harvard.edu/abs/2018MNRAS.477.2804S} {477, 2804}

\bibitem[\protect\citeauthoryear{{Springel}}{{Springel}}{2005}]{2005MNRAS.364.1105S}
{Springel} V.,  2005, \mn@doi [\mnras] {10.1111/j.1365-2966.2005.09655.x},
  \href {http://adsabs.harvard.edu/abs/2005MNRAS.364.1105S} {364, 1105}

\bibitem[\protect\citeauthoryear{{Springel}, {White}, {Tormen}  \&
  {Kauffmann}}{{Springel} et~al.}{2001}]{2001MNRAS.328..726S}
{Springel} V.,  {White} S.~D.~M.,  {Tormen} G.,   {Kauffmann} G.,  2001,
  \mn@doi [\mnras] {10.1046/j.1365-8711.2001.04912.x}, \href
  {http://adsabs.harvard.edu/abs/2001MNRAS.328..726S} {328, 726}

\bibitem[\protect\citeauthoryear{{Springel}, {Di Matteo}  \&
  {Hernquist}}{{Springel} et~al.}{2005a}]{2005MNRAS.361..776S}
{Springel} V.,  {Di Matteo} T.,   {Hernquist} L.,  2005a, \mn@doi [\mnras]
  {10.1111/j.1365-2966.2005.09238.x}, \href
  {http://adsabs.harvard.edu/abs/2005MNRAS.361..776S} {361, 776}

\bibitem[\protect\citeauthoryear{{Springel} et~al.,}{{Springel}
  et~al.}{2005b}]{2005Natur.435..629S}
{Springel} V.,  et~al., 2005b, \mn@doi [\nat] {10.1038/nature03597}, \href
  {http://adsabs.harvard.edu/abs/2005Natur.435..629S} {435, 629}

\bibitem[\protect\citeauthoryear{{Steinborn}, {Dolag}, {Hirschmann}, {Prieto}
  \& {Remus}}{{Steinborn} et~al.}{2015}]{2015MNRAS.448.1504S}
{Steinborn} L.~K.,  {Dolag} K.,  {Hirschmann} M.,  {Prieto} M.~A.,   {Remus}
  R.-S.,  2015, \mn@doi [\mnras] {10.1093/mnras/stv072}, \href
  {http://adsabs.harvard.edu/abs/2015MNRAS.448.1504S} {448, 1504}

\bibitem[\protect\citeauthoryear{{Steinborn}, {Dolag}, {Comerford},
  {Hirschmann}, {Remus}  \& {Teklu}}{{Steinborn}
  et~al.}{2016}]{2016MNRAS.458.1013S}
{Steinborn} L.~K.,  {Dolag} K.,  {Comerford} J.~M.,  {Hirschmann} M.,  {Remus}
  R.-S.,   {Teklu} A.~F.,  2016, \mn@doi [\mnras] {10.1093/mnras/stw316}, \href
  {http://adsabs.harvard.edu/abs/2016MNRAS.458.1013S} {458, 1013}

\bibitem[\protect\citeauthoryear{{Teklu}, {Remus}, {Dolag}, {Beck}, {Burkert},
  {Schmidt}, {Schulze}  \& {Steinborn}}{{Teklu}
  et~al.}{2015}]{2015ApJ...812...29T}
{Teklu} A.~F.,  {Remus} R.-S.,  {Dolag} K.,  {Beck} A.~M.,  {Burkert} A.,
  {Schmidt} A.~S.,  {Schulze} F.,   {Steinborn} L.~K.,  2015, \mn@doi [\apj]
  {10.1088/0004-637X/812/1/29}, \href
  {http://adsabs.harvard.edu/abs/2015ApJ...812...29T} {812, 29}

\bibitem[\protect\citeauthoryear{{Teklu}, {Remus}  \& {Dolag}}{{Teklu}
  et~al.}{2016}]{2016ilgp.confE..41T}
{Teklu} A.~F.,  {Remus} R.-S.,   {Dolag} K.,  2016, in The Interplay between
  Local and Global Processes in Galaxies,. p.~41

\bibitem[\protect\citeauthoryear{{Tollet} et~al.,}{{Tollet}
  et~al.}{2016}]{2016MNRAS.456.3542T}
{Tollet} E.,  et~al., 2016, \mn@doi [\mnras] {10.1093/mnras/stv2856}, \href
  {http://adsabs.harvard.edu/abs/2016MNRAS.456.3542T} {456, 3542}

\bibitem[\protect\citeauthoryear{{Umetsu}, {Zitrin}, {Gruen}, {Merten},
  {Donahue}  \& {Postman}}{{Umetsu} et~al.}{2016}]{2016ApJ...821..116U}
{Umetsu} K.,  {Zitrin} A.,  {Gruen} D.,  {Merten} J.,  {Donahue} M.,
  {Postman} M.,  2016, \mn@doi [\apj] {10.3847/0004-637X/821/2/116}, \href
  {http://adsabs.harvard.edu/abs/2016ApJ...821..116U} {821, 116}

\bibitem[\protect\citeauthoryear{{Voevodkin}, {Borozdin}, {Heitmann}, {Habib},
  {Vikhlinin}, {Mescheryakov}, {Hornstrup}  \& {Burenin}}{{Voevodkin}
  et~al.}{2010}]{2010ApJ...708.1376V}
{Voevodkin} A.,  {Borozdin} K.,  {Heitmann} K.,  {Habib} S.,  {Vikhlinin} A.,
  {Mescheryakov} A.,  {Hornstrup} A.,   {Burenin} R.,  2010, \mn@doi [\apj]
  {10.1088/0004-637X/708/2/1376}, \href
  {http://adsabs.harvard.edu/abs/2010ApJ...708.1376V} {708, 1376}

\bibitem[\protect\citeauthoryear{{Wang}, {Dutton}, {Stinson}, {Macci{\`o}},
  {Penzo}, {Kang}, {Keller}  \& {Wadsley}}{{Wang}
  et~al.}{2015}]{2015MNRAS.454...83W}
{Wang} L.,  {Dutton} A.~A.,  {Stinson} G.~S.,  {Macci{\`o}} A.~V.,  {Penzo} C.,
   {Kang} X.,  {Keller} B.~W.,   {Wadsley} J.,  2015, \mn@doi [\mnras]
  {10.1093/mnras/stv1937}, \href
  {http://adsabs.harvard.edu/abs/2015MNRAS.454...83W} {454, 83}

\bibitem[\protect\citeauthoryear{{Zhao}, {Jing}, {Mo}  \& {B{\"o}rner}}{{Zhao}
  et~al.}{2009}]{2009ApJ...707..354Z}
{Zhao} D.~H.,  {Jing} Y.~P.,  {Mo} H.~J.,   {B{\"o}rner} G.,  2009, \mn@doi
  [\apj] {10.1088/0004-637X/707/1/354}, \href
  {http://adsabs.harvard.edu/abs/2009ApJ...707..354Z} {707, 354}

\bibitem[\protect\citeauthoryear{{van de Sande} et~al.,}{{van de Sande}
  et~al.}{2019}]{2019MNRAS.484..869V}
{van de Sande} J.,  et~al., 2019, \mn@doi [\mnras] {10.1093/mnras/sty3506},
  \href {http://adsabs.harvard.edu/abs/2019MNRAS.484..869V} {484, 869}

\makeatother
\end{thebibliography}

\bsp	
\label{lastpage}
\end{document}